\documentclass[twocolumn,preprintnumbers,superscriptaddress,nofootinbib,aps,prd,floatfix]{revtex4-2}
\pdfoutput=1
\usepackage{enumerate}
\usepackage{amsmath,amssymb}
\usepackage{graphicx}
\usepackage{slashed}
\usepackage{float}
\usepackage[dvipsnames]{xcolor}
\usepackage[normalem]{ulem}
\usepackage{subfigure,orcidlink}
\usepackage{multirow,array}
\usepackage{hyperref}
\hypersetup{%
	colorlinks = true,%
	linkcolor = Blue,%
	citecolor = Blue,%
	filecolor = Blue,%
	urlcolor = Blue%
}
\usepackage{tabularray}
\UseTblrLibrary{booktabs}

\begin{document}
\title{Thermal leptogenesis, dark matter, and gravitational waves\\from an extended canonical seesaw scenario}

\author{Partha Kumar Paul\orcidlink{https://orcid.org/0000-0002-9107-5635}}
\email{ph22resch11012@iith.ac.in}
\affiliation{Department of Physics, Indian Institute of Technology Hyderabad, Kandi, Telangana-502285, India.}

\author{Narendra Sahu\orcidlink{https://orcid.org/0000-0002-9675-0484}}
\email{nsahu@phy.iith.ac.in}
\affiliation{Department of Physics, Indian Institute of Technology Hyderabad, Kandi, Telangana-502285, India.}

\author{Prashant Shukla\orcidlink{https://orcid.org/0000-0001-8118-5331}}
\email{pshukla@barc.gov.in}
\affiliation{Nuclear Physics Division, Bhabha Atomic Research Centre,
	Mumbai, 400085, India.}
\affiliation{Homi Bhabha National Institute, Anushakti Nagar, Mumbai,
	400094, India.}
	
\date{\today}

	\begin{abstract}
	
	In a canonical type-I seesaw scenario, the Standard Model is extended with three singlet right-handed neutrinos (RHNs) $N_i, i=1,2,3$ 
with masses $M_i, i=1,2,3$ to simultaneously explain sub-eV masses of light neutrinos and baryon asymmetry of the Universe at high scales. In this paper, we show 
that a relatively low-scale thermal leptogenesis accompanied by gravitational wave signatures is possible when the type-I seesaw is extended with 
a singlet fermion ($S$) and a singlet scalar ($\rho$), where $S$ and $\rho$ are odd under a discrete $Z_2$ symmetry. We also add a vectorlike fermion 
doublet $\Psi$ and impose a $Z^\prime_2$ symmetry under which both $N_1$ and $\Psi$ are odd while all other particles are even. This gives rise to a 
singlet-doublet Majorana fermion dark matter in our setup. At a high scale, the  $Z_2$ symmetry is broken spontaneously by the vacuum expectation 
value of $\rho$ and leads to (i) mixing between  RHNs ($N_2, N_3$) and $S$, and (ii) formation of Domain walls (DWs). In the former case, the final lepton 
asymmetry is generated by the out-of-equilibrium decay of $S$, which dominantly mixes with $N_2$. We show that the scale of thermal leptogenesis can be 
lowered to $M_S \sim 2 \times 10^6$ GeV, which is \textit{3} orders of magnitude lower than the thermal leptogenesis in canonical type-I seesaw. In the latter case, the disappearance of the DWs gives observable gravitational wave signatures, which can be 
probed at LISA, DECIGO, ${\rm \mu ARES}$ etc.

\end{abstract}
\maketitle

\noindent
\section{Introduction}
\label{intro}

Recently NANOGrav\cite{NANOGrav:2023gor,NANOGrav:2023hvm}, EPTA \cite{EPTA:2023fyk},  PPTA \cite{Reardon:2023gzh} have reported positive evidence for stochastic gravitational wave (GW) background in the nanohertz (nHz) frequency range. This stochastic GW background could originate from various sources. One such possibility is the domain walls (DWs), which are two-dimensional topological defects that emerge when a discrete symmetry is spontaneously broken in the early Universe \cite{Vilenkin:2000jqa}. In cosmology, the formation of DWs possess a significant challenge because their energy density can quickly surpass the total energy density of the Universe, which contradicts current observational data \cite{Zeldovich:1974uw}. However, it's possible that DWs are inherently unstable and collapse before they can dominate the total energy density of the Universe. This instability could be due to an explicit breaking of the discrete symmetry in the underlying theory \cite{Vilenkin:1981zs,Gelmini:1988sf,Larsson:1996sp}. If this is the case, a considerable amount of GWs could be generated during the collisions and annihilations of these domain walls \cite{Gleiser:1998na,Hiramatsu:2010yz,Kawasaki:2011vv,Hiramatsu:2013qaa,Bhattacharya:2023kws,Borah:2022wdy,Barman:2022yos}. These GWs might persist as a stochastic background in the universe today. Detecting these GWs would offer insights into early cosmic events and provide a novel method for exploring physics at extremely high energies. 

\begin{figure}[]
	\includegraphics[scale=0.13]{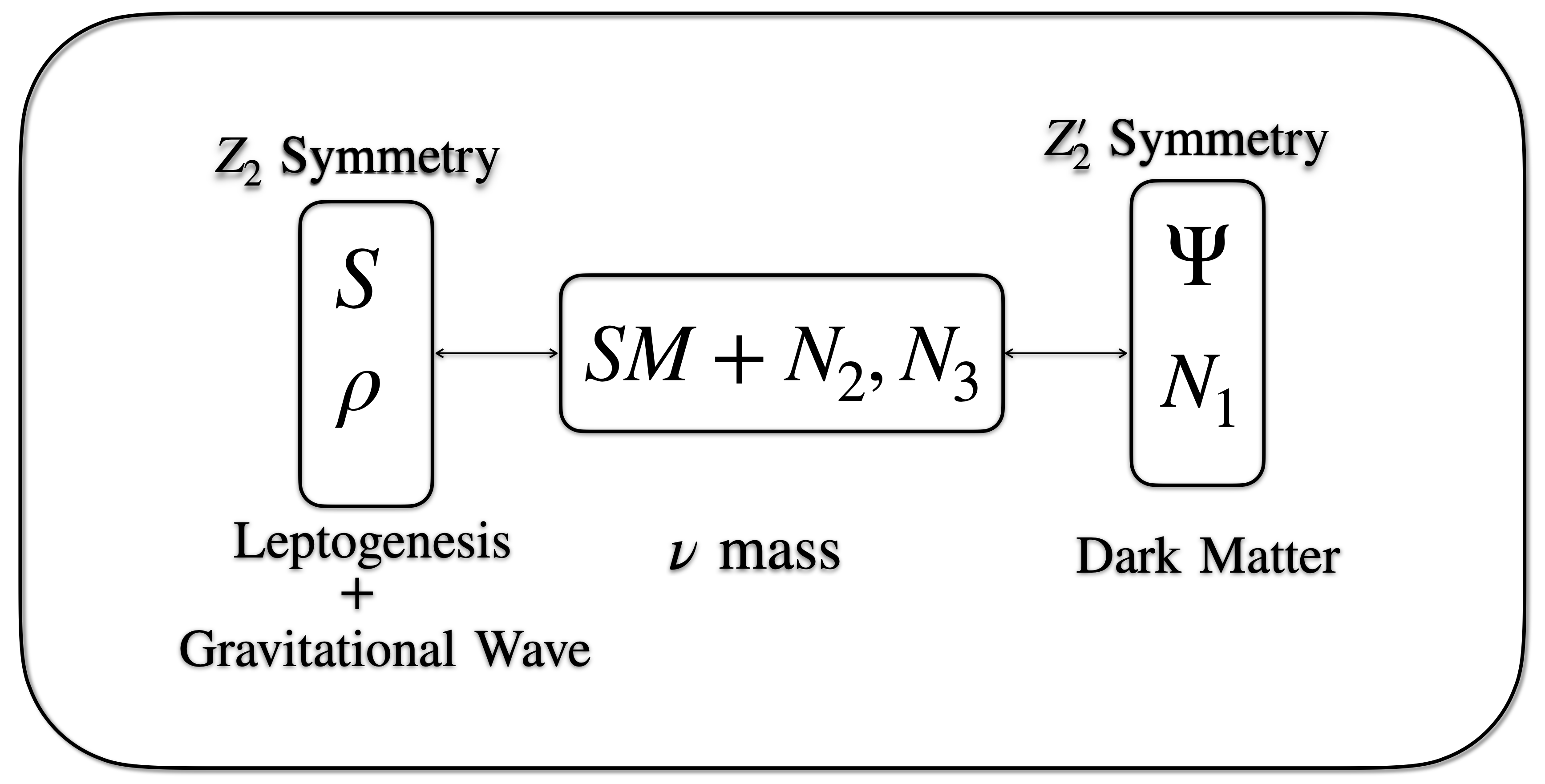}
	\caption{The schematic picture of our setup.}\label{fig:pic}
\end{figure}

In this paper, we try to connect the nonzero neutrino mass, dark matter (DM), observed baryon asymmetry of the Universe and the GW sourced by the DWs in a common framework as shown pictorially in Fig. \ref{fig:pic}.  With this motivation, we extend the canonical type-I seesaw model\cite{Minkowski:1977sc,Mohapatra:1980yp,Mohapatra:1991ng,Schechter:1980gr,Valle:2015pba} with a singlet fermion $S$ and one singlet scalar $\rho$. We impose a  discrete $Z_2$ symmetry under which $S$ and $\rho$ are odd while all other particles are even.  This forbids the direct coupling of $S$ with the $L,H$. We also introduce a vectorlike fermion doublet $\Psi \equiv \left( \psi^0~ \psi^-\right)^T$ and impose a discrete symmetry $Z^\prime_2$ under which $\Psi$ and the lightest RHN ($N_1$) are odd while all other particles are even. 
This gives rise to a singlet-doublet Majorana fermion DM~\cite{Mahbubani:2005pt,DEramo:2007anh,Cohen:2011ec,Freitas:2015hsa,Cynolter:2015sua,Calibbi:2015nha,Cheung:2013dua,Enberg:2007rp,Banerjee:2016hsk,DuttaBanik:2018emv,Horiuchi:2016tqw,Restrepo:2015ura,Abe:2017glm,Bhattacharya:2017sml,Bhattacharya:2018cgx,Bhattacharya:2016rqj,Bhattacharya:2015qpa,Bhattacharya:2018fus,Bhattacharya:2021ltd,Dutta:2020xwn,Borah:2021khc,Borah:2021rbx,Borah:2022zim,Borah:2023dhk}.  At a high scale, when the singlet scalar $\rho$ obtains a vacuum expectation value (vev), the $Z_2$ symmetry gets broken and leads to (i) mixing between the singlet fermion $S$ and RHNs ($N_2,N_3$) and (ii) formation of Domain walls (DWs) \cite{Zeldovich:1974uw,Kibble:1976sj,Vilenkin:1981zs,Kibble:1982dd,Lazarides:1981fv,Vilenkin:1984ib}. In the former case, the $S$ decays to the Standard Model (SM) lepton, $L$ and $H$ via the mixing with $N_{2,3}$. Assuming a hierarchy among the $S$ and $N_{2,3}$, the final lepton asymmetry is established by the 
\textit{CP} violating out-of-equilibrium decay of $S$. Due to the mixing, the decay width of $S\rightarrow LH$ will be suppressed compared to the usual type-I case $N\rightarrow LH$. As a result, the out-of-equilibrium condition can be satisfied for lower values of the mass of $S$. The \textit{CP} asymmetry parameter does not have any suppression. It depends on the value of $M_S$ unlike $M_{N}$ in the usual case. However, the value of $M_S$ cannot be arbitrarily small since it will make the \textit{CP} asymmetry parameter $\epsilon_S$ very small. We find that the leptogenesis scale can be brought down to $\mathcal{O}(10^6)$ GeV, which is \textit{3} orders of magnitude smaller than the thermal leptogenesis in canonical type-I seesaw. For low scale leptogenesis in various beyond Standard Model frameworks, see,  \textit{e.g.}, incorporating flavor effects \cite{Blanchet:2008pw}, by adding
	extra scalar fields \cite{Clarke:2015hta,Hugle:2018qbw,Borah:2021qmi,Vatsyayan:2022rth}, resonant leptogenesis \cite{Pilaftsis:2003gt}, right-handed sector leptogenesis \cite{Frigerio:2006gx,McDonald:2007ka}, wash-in leptogenesis \cite{Domcke:2020quw}, and by decoupling neutrino mass and leptogenesis \cite{Ma:2006te,Ma:2006ci,Hirsch:2006ft,Hambye:2006zn}.
	
	Another intriguing consequence of the discrete $Z_2$ symmetry breaking is the formation of DWs. We analyze the dynamics of these DWs. To ensure their instability, we introduce an explicit $Z_2$-breaking term in the scalar potential, creating a pressure difference across the walls. This causes the DWs to collapse and annihilate, leading to the production of stochastic GWs. Crucially, in our framework, it is the vev of $\rho$ that establishes the connection between leptogenesis and GW signals. We identify the parameter space where both leptogenesis and GW production occur simultaneously.

The paper is organized as follows. We discuss the model in Sec. \ref{sec:model}.  In Sec. \ref{sec:singletlepto} we discuss 
a low energy thermal leptogenesis scenario arising from a $Z_2$ symmetry breaking. The domain wall dynamics and signature of gravitational waves are discussed in Sec. \ref{sec:dws}. The DM phenomenology is discussed in Sec. \ref{sec:dmpheno}. We finally conclude in Sec. \ref{sec:conclusion}.

\section{The model}\label{sec:model}
 We extend the SM with three RHNs, $N_i, i=1,2,3$ with masses $M_i, i=1,2,3$, one singlet fermion $S$ with mass $M_S$ and one singlet scalar $\rho$. 
 
 \begin{table}[h]
 	\centering
 	\caption{Particles and their charge assignment under the imposed symmetry.}
 	\resizebox{7cm}{!}{
 		\begin{tblr}{
 				colspec={llllllll},
 				row{1}={font=\bfseries},
 				column{1}={font=\itshape},
 				row{1}={bg=gray!40},row{2}={bg=gray!30},row{3}={bg=gray!25},row{4}={bg=gray!20},row{5}={bg=gray!15},row{6}={bg=gray!10}
 			}
 			\toprule {\rm Field}&$SU(3)_c$&$SU(2)_L$&$U(1)_Y$&$Z_2$&$Z_2^\prime$\\
 			\toprule
 			$N_1$&
 			1&1& 0&+&-- \\
 			
 		$N_{2,3}$&
 		1&1& 0&+&+ \\
 			
 			$S$&
 			1&1& 0&--&+ \\
 			$\rho$&
 			1&1& 0&--&+ \\
$\Psi$&
1&2& --1&+&-- \\
 			\bottomrule
 	\end{tblr}}
 	\label{tab:tab1}
 \end{table}
We impose a discrete symmetry $Z_2$ under which $S$ and $\rho$ are odd, while all other particles are even. As a 
result the direct coupling of $S$ with $L,H$ is forbidden. We also introduce a vectorlike fermion doublet $\Psi \equiv \left( \psi^0~ \psi^-\right)^T$ and impose an additional discrete symmetry $Z^\prime_2$ under which $\Psi$ and $N_1$ are odd, while 
all other particles are even as listed in Table \ref{tab:tab1}\footnote{This restricts the $N_1 L H$ coupling, thus resulting in only two nonzero SM light neutrino masses.}. The neutral component of $\Psi$ combines with $N_1$ to give rise to a singlet-doublet fermion dark matter. 
  The relevant Lagrangian involving $N$, $S$ and $\rho$ is given as,
\begin{eqnarray}
	\mathcal{L}&=&\bar{N}i\gamma_\mu\partial^\mu N+\bar{S}i\gamma_\mu\partial^\mu S-y_{_{N l}}\bar{L}\tilde{H}N-y_{_{N S}}\bar{N}\rho S\nonumber\\&-&\frac{1}{2}M_S \bar{S^c}S-\frac{1}{2}M_{N}\bar{N^c}N+{\rm H.c.},
\end{eqnarray}
where $N$ represents $N_2,N_3$.

The scalar potential involving the SM Higgs doublet $H$ and singlet scalar $\rho$ can be written as
\begin{eqnarray}
	V(H,\rho)&=&	-\mu_H^2 H^\dagger H-\mu^2_\rho \rho^2 +\lambda_H(H^\dagger H)^2+\lambda_\rho \rho^4\nonumber\\&+&\lambda_{H\rho}(H^\dagger H)\rho^2.
\end{eqnarray}
Due to negative mass-squared term of $\rho$, the latter acquires a vev: $\langle \rho\rangle\neq0$ which leads to the spontaneous breaking of $Z_2$ symmetry. The minimum of the potential is at $\langle\rho\rangle=v_\rho=\pm\sqrt{\frac{\mu_\rho^2}{\lambda_\rho}}$. 
	Here $\mu_\rho=\sqrt{\lambda_\rho v_\rho^2}$.
	At finite temperature, interaction of the scalar field $\rho$ with others fields may induce a thermal correction to the potential which leads to thermal mass of $\rho$. In our model, $\rho$ interacts with other fermions via the Yukawa interaction term $y_{NS}NS\rho$.
	This leads to an effective potential of the form \cite{Senaha:2020mop,Matsedonskyi:2020mlz}
	\begin{eqnarray}
		V_{\rm eff}(\rho)&=&-\mu^2_\rho \rho^2 +\lambda_\rho \rho^4+cT^2\rho^2\nonumber\\&=&(-\mu^2_\rho+cT^2) \rho^2 +\lambda_\rho \rho^4,
	\end{eqnarray}
	where $c=\frac{y_{NS}^2} {4}+\frac{\lambda_{H\rho}}{12}+\frac{\lambda_\rho}{2}$. At high temperature the $cT^2$ term dominates over the $-\mu_\rho^2$ term and the sign of the effective mass term changes when 
	\begin{eqnarray}
		m_{\rm eff}^2(T)=-\mu_{\rho}^2+cT^2>0.
	\end{eqnarray}
	This leads to a critical temperature for symmetry restoration which is given by
	\begin{eqnarray}
		T_C=\sqrt{\frac{\mu_\rho^2}{c}}
	\end{eqnarray}
	Thus for $T>T_C$, the quadratic coefficient of the field becomes positive, which sets $v_{\rho}=0$ and restores the $Z_2$ symmetry. For $T<T_C$, the $v_\rho\neq0$ and the symmetry is broken. The vev of the $\rho$ is then given at different temperatures as
	\begin{equation}
		v_\rho(T)=\left\{
		\begin{array}{l}
			0~~~~~~~~~~~~~~~~~~~~~~~~~~~~~T\geq T_C\\
			\sqrt{\frac{\mu_\rho^2-cT^2}{\lambda_\rho}}~~~~~~~~~~~~~~~~~~T<T_C\\
			\sqrt{\frac{\mu_\rho^2}{\lambda_\rho}}\equiv v_\rho~~~~~~~~~~~~~~~~~T=0.\\	\end{array}
		\right.
	\end{equation}
	Therefore, the mass of the $\rho$  at different temperatures can be given  as
	\begin{equation}
		M_\rho(T)=\left\{
		\begin{array}{l}
			0~~~~~~~~~~~~~~~~~~~~~~~~~~~~T\geq T_C\\
			\sqrt{2(\mu_\rho^2-cT^2)}~~~~~~~~~~~T<T_C\\
			\sqrt{2}\mu_\rho~~~~~~~~~~~~~~~~~~~~~~T=0.\\	\end{array}
		\right.
	\end{equation}
The electroweak symmetry breaking (EWSB) occurs when the SM Higgs obtains a vev $v_h$. The vacuum fluctuations of the scalar fields at zero temperature  are given as
\begin{eqnarray}
	H=\frac{1}{\sqrt{2}}(0,h+v_h)^T, ~~ \rho=\frac{\rho^\prime+v_\rho}{\sqrt{2}},
\end{eqnarray}
where the vevs are $v_h=\sqrt{\frac{\mu_H^2}{\lambda_H}}$, and $v_\rho=\sqrt{\frac{\mu_\rho^2}{\lambda_\rho}}$.

\subsection{Neutrino mass}\label{sec:neutrinomass}
Above electroweak phase transition (EWPT), the singlet scalar $\rho$ acquires a vev: $v_\rho$ and breaks the $Z_2$ symmetry spontaneously. This allows the singlet fermion $S$ to mix with the RHNs $N_2$ and $N_3$. In the effective theory, the $6\times 6$ 
fermion mass matrix can be written in the basis $[L_i,N_{2,3},S]$ as
\begin{eqnarray}
\mathcal{M}=\begin{pmatrix}
0&m_{D}&0\\
m_{D}&M_{N} & d\\
0&d& M_S
\end{pmatrix},
\end{eqnarray}
where $d=y_{_{N S}}v_{\rho}/\sqrt{2}$, $m_{D}=y_{_{N l}}v_h/\sqrt{2}$.

Diagonalizing this mass matrix we obtain the masses of heavy eigenstates as
\begin{eqnarray}
M_N^\prime\simeq M_{N}+\frac{d^2}{M_{N}-M_S},\\
M_{S}^\prime\simeq M_{S}-\frac{d^2}{M_N-M_S},
\end{eqnarray}
where the mixing angle is given as
\begin{eqnarray}
\theta\simeq \frac{d}{M_{N}-M_S}.
\end{eqnarray}
The light neutrino mass matrix is then given as,
\begin{eqnarray}
(m_{\nu})_{ij}\simeq-\sum_k(m_D)_{ik}\bigg( M_{k}+\frac{d^2}{M_k-M_S} \bigg)^{-1}(m_{D})_{kj}.\nonumber\\
\end{eqnarray}
The canonical type-I seesaw can be restored in the limit $d\rightarrow0$.


\section{Thermal leptogenesis from $Z_2$ symmetry breaking}\label{sec:singletlepto}
\begin{figure}[h]
	\includegraphics[scale=0.7]{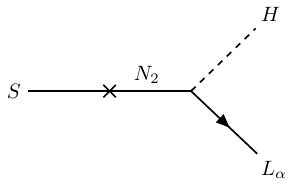}
	\includegraphics[scale=0.7]{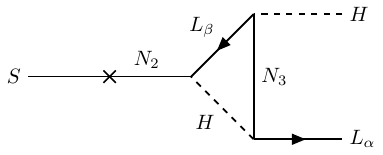}
	\includegraphics[scale=0.7]{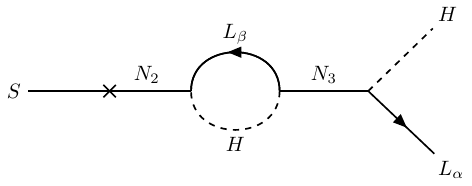}
	\caption{Tree and one loop diagrams of $S$ decay giving rise to \textit{CP} asymmetry.}\label{fig:CPdiag}
\end{figure}

\begin{figure}[h]
	\includegraphics[scale=0.35]{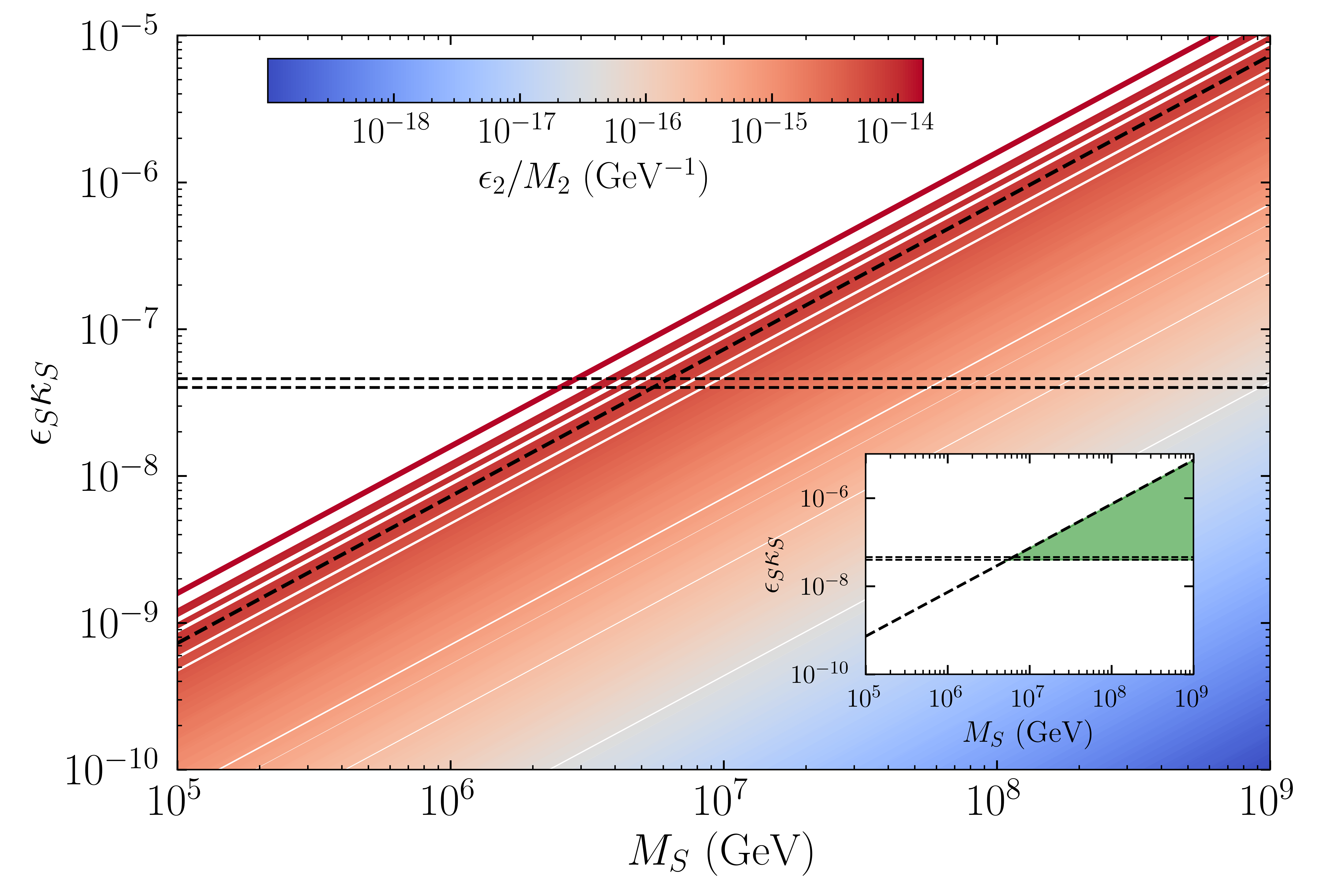}
	\caption{Values of $\epsilon_S\kappa_S$ as a function of $M_S$. Each of the lines corresponds to a set of parameters when $\epsilon_{2}\kappa_2\leq 10^{-9}$. The color code represents the values of $\epsilon_{2}/M_2$. The dashed horizontal lines represent the observed baryon asymmetry, which translates to $\epsilon_S\kappa_S=(4.38\pm0.22)\times10^{-8}$. The lowest allowed value of $M_S$ corresponds to $\epsilon_{2}/M_2\sim2\times10^{-14}\rm GeV^{-1}$.  }\label{fig:summary}
\end{figure}

In the type-I seesaw, the \textit{CP} violating out-of-equilibrium decay of heavy neutrinos to the SM sector (leptons and Higgs) can generate a net lepton asymmetry which can be converted to the baryon asymmetry via the electroweak sphalerons. In our scenario, the lepton asymmetry can be generated from the direct decay of $N_2,N_3$ as well as from the decay of $S$ via its mixing with $N_2$, where we assume the mixing with $N_3$ is negligible. We are interested in the region where the final asymmetry is produced mainly due to mixing suppressed decay of $S$. However, while calculating the lepton asymmetry, we simultaneously take into account the effects of $N_2$ and $N_3$ along with $S$. We have also taken into account the scale dependence of the coupling by solving the renormalization group (RG) equations as given in Appendix \ref{app:rgeq}. We used \texttt{SARAH 4.13.0} \cite{Staub:2013tta} to calculate the RG equations.  The values of the couplings listed in Table \ref{tab:tab2} are taken at the  EW scale. The decay width of the $N_{2,3}$ is given by
\begin{eqnarray}
	\Gamma_i=\frac{(y^\dagger_{Nl}y_{Nl})_{ii}}{8\pi}M_i,
\end{eqnarray}
where $M_i,i=2,3$ are the masses of $N_2$ and $N_3$ respectively.
The decay width of the singlet fermion  $S$ is then given by
\begin{eqnarray}
	\Gamma_S=\theta_s(T)^2\frac{(y^\dagger_{_{N l}}y_{_{N l}})_{22}}{8\pi}M_{S},\label{eq:gammaS1}
\end{eqnarray}
where $\theta_s(T)$ represents the mixing between $S-N_2$ at finite temperature and is given by
\begin{eqnarray}
	\theta_s(T)=\frac{y_{_{N S}}v_{\rho}(T)}{\sqrt{2}(M_{2}-M_S)}.\label{eq:theta1}
\end{eqnarray}
The Yukawa coupling, $y_{_{N l}}$	in Eq \ref{eq:gammaS1} can be calculated by using the Casas-Ibarra parametrization as \cite{Casas:2001sr}
\begin{eqnarray}
	y_{_{N l}}=i\frac{\sqrt{2}}{v_h}(U^*_{\rm PMNS}\cdot\sqrt{\hat{m}_\nu}\cdot R^T\cdot\sqrt{\hat{M}_N}), \label{eq:CI}
\end{eqnarray}
where $U_{\rm PMNS}$ is the lepton mixing matrix, $\hat{m}_\nu$ is $3\times3$ diagonal light neutrino mass matrix with
eigenvalues $0,m_2$, and $m_3$ ; $\hat{M}_N$ is $3\times3$ diagonal RHN mass matrix with eigenvalues
$0, M_{2}$, and $M_3$; $R$ is an arbitrary rotation matrix.\\
The \textit{CP} asymmetry parameters from the decay of $N_{2,3}\rightarrow LH$ arises via the interference of the tree and one loop diagrams and this is given by  \cite{Ma:2006te,Ma:2006ci}
\begin{eqnarray}
	\epsilon_{i}=-\frac{1}{8\pi}\frac{1}{(y^\dagger_{Nl}y_{Nl})_{ii}}\sum_{j\neq i}\text{Im}\{(y^\dagger_{Nl}y_{Nl})^2_{ij}\}\bigg[f_v\bigg(\frac{M_j^2}{M_i^2}\bigg)+ f_s\bigg(\frac{M_j^2}{M_i^2}\bigg) \bigg],\nonumber\\\label{eq:cpasymm}
\end{eqnarray}
where the $f_v(x),f_s(x)$ denote the one loop vertex and self-energy corrections given as
\begin{eqnarray}
	f_v(x)&=&\sqrt{x}\big[1-(1+x)\ln\big(\frac{x}{1+x}\big)\big];f_s(x)=\frac{\sqrt{x}}{1-x}.
\end{eqnarray}
In a hierarchical limit, the \textit{CP} asymmetry parameters can be estimated from Eq \ref{eq:cpasymm} as
\begin{eqnarray}
	\epsilon_i=-\frac{3}{8\pi v_h^2}\sum_{j\neq i}\frac{M_{i}}{M_j}\frac{Im[(m^\dagger_Dm_D)_{ij}]^2}{(m^\dagger_Dm_D)_{ii}}
\end{eqnarray}
The \textit{CP} asymmetry ($\epsilon_S$) generated by the decay of $S$ which comes from the interference of the tree and one loop diagrams is shown in Fig \ref{fig:CPdiag} and can be expressed as
\begin{eqnarray}
	\epsilon_S=-\frac{3}{8\pi v_h^2}\frac{M_{S}}{M_3}\frac{Im[(m^\dagger_Dm_D)_{23}]^2}{(m^\dagger_Dm_D)_{22}}.\label{eq:CPasy1}
\end{eqnarray}
Thus the \textit{CP} asymmetry parameters of $S$ and $N_2$ can be related as
\begin{eqnarray}
	\epsilon_S=(M_S/M_2)\epsilon_2.\label{eq:epsilonS}
\end{eqnarray}
From Eq \ref{eq:epsilonS}, we see that $M_S$ cannot be arbitrarily small in order to produce correct lepton asymmetry via the decay of $S$ through $S-N_2$ mixing. Above electroweak phase transition, the net lepton asymmetry produced by $N_2,N_3$, and $S$ can be converted to the observed baryon asymmetry via nonperturbative sphaleron processes. The resulting baryon asymmetry can be given as,	
\begin{eqnarray}
	\eta_{_ B}(z\rightarrow\infty)&=&\frac{C_{L\rightarrow B}}{f}\sum_i Y_{\Delta L}(z\rightarrow\infty)\nonumber\\&=&\frac{C_{L\rightarrow B}}{f}\sum_i\epsilon_i \kappa_i(z\rightarrow\infty) Y^{\rm eq}_i(z\rightarrow0)\nonumber\\&=&-0.01394\sum_i\epsilon_i \kappa_i(z\rightarrow\infty),\nonumber\\\label{eq:etab}
\end{eqnarray}
where, $i\in\{N_2,N_3,S\}$, $z=M_S/T$, $\kappa_i$ is the efficiency factor, $Y^{\rm eq}_i$ is the equilibrium abundance of $i$th particle defined as, $Y^{\rm eq}_i=n^{\rm eq}_i/n_\gamma$, $n^{\rm eq}_i$ is the equilibrium number density of $i$th species, $n_\gamma$ is the photon number density, $f=29.54${\footnote{$g^*_s=106.75+\frac{7}{8}\times2+\frac{7}{8}\times(4+4)=115.5$, is the relativistic d.o.f at the onset of leptogenesis and $g^*_0=3.91$ is the relativistic d.o.f today. The dilution factor $f$ is then , $f=g^*_s/g^*_0=29.54$.}} is the dilution factor calculated assuming standard photon production from the onset of leptogenesis
till recombination, $C_{L\rightarrow B}$ is the lepton to baryon asymmetry conversion factor and in our case the value of it is $C_{L\rightarrow B}=-0.54902$\footnote{$C_{L\rightarrow B}=-\frac{4m+8n}{9m+14m}$. For the steps of the calculation refer to \cite{Mahapatra:2023dbr,Borah:2024wos}.}. We note that $Y^{\rm eq}_i(z\rightarrow0)=3/4$. The required value of the baryon asymmetry is $\eta_{_B}=(6.1\pm0.3)\times10^{-10}$ \cite{Planck:2018vyg}, which translates to the required lepton asymmetry as $\sum_i Y_{\Delta L}\equiv \sum_i\epsilon_i \kappa_i Y_i^{\rm eq}(z\rightarrow0)\sim\mathcal{O}( 10^{-8})$. As discussed above the lepton asymmetry production due to $S$ decay is via its mixing with $N_2$. Therefore, if the out-of-equilibrium decay of $S$ happens at late epoch, the washout processes will be no more active. In that case, the efficiency factor, $\kappa_S$ will be $\sim 1$. From Eq \ref{eq:etab}, we see that $S$ will contribute to the lepton asymmetry significantly, when the contributions from $N_2,N_3$ are negligible. This can be realized only if $\epsilon_{2}\kappa_{2},\epsilon_{3}\kappa_{3}\ll 10^{-8}$\cite{Engelhard:2006yg,Paul:2025iks}. In that case, Eq \ref{eq:etab} will be reduced to
\begin{eqnarray}
	\eta_{_ B}=-0.01394\epsilon_S \kappa_S.\label{eq:etabS}
\end{eqnarray}
Thus from Eq \ref{eq:cpasymm} and \ref{eq:etabS} we get a lower limit on the $M_S$ when the $\kappa_S\sim1$ (\textit{i.e.} $S$ is in weak washout regime which is true due to the late decay of $S$ when all the washouts processes are suppressed) to be
\begin{eqnarray}
	M_{S}\gtrsim2.15\times10^6{\rm GeV}\bigg(\frac{\eta_B}{6\times10^{-10}}\bigg)\bigg(\frac{2\times10^{-14}{\rm GeV^{-1}}}{\epsilon_2/M_2}\bigg),\nonumber\\ \label{eq:msbound}
\end{eqnarray}
where the conservative upper bound on $\epsilon_{2}/M_2$ is obtained by choosing $\epsilon_2\kappa_2\leq10^{-9}$. The details are given in Fig \ref{fig:summary}.\\

To see the impact of $N_2$ and $N_3$ on the final lepton asymmetry, we solve the complete set of BEs for the evolution of lepton asymmetry as well as the abundances of $S,N_2$, and $N_3$ which are given as
{\scriptsize\begin{eqnarray}
	\frac{dY_{N_2}}{dz}&=&-\frac{\Gamma^{D}_2}{\mathcal{H}z}(Y_{N_2}-Y_{N_2}^{\rm eq})-\frac{\Gamma^\prime_{1}}{\mathcal{H}z}(Y_{N_2}-Y_{N_2}^{\rm eq}),\label{eq:dYn2}
\end{eqnarray}
\begin{eqnarray}
	\frac{dY_{N_3}}{dz}&=&-\frac{\Gamma^{D}_3}{\mathcal{H}z}(Y_{N_3}-Y_{N_3}^{\rm eq})-\frac{\Gamma^{\prime\prime}_{1}}{\mathcal{H}z}(Y_{N_3}-Y_{N_3}^{\rm eq}),\label{eq:dYn3}
\end{eqnarray}
\begin{eqnarray}
	\frac{dY_S}{dz}&=&-\frac{\Gamma^D_S}{\mathcal{H}z}(Y_S-Y_S^{\rm eq})-\frac{(\Gamma_{1}+\Gamma_{0})}{\mathcal{H}z}(Y_S-Y_S^{\rm eq})-\frac{\Gamma^\prime_{0}}{\mathcal{H}z}\frac{(Y^2_S-(Y_S^{\rm eq})^2)}{Y_S^{\rm eq}},\label{eq:dYs2}\nonumber\\
\end{eqnarray}
\begin{eqnarray}
	\frac{dY^{\rm all}_{\Delta L}}{dz}&=&\epsilon_S\frac{\Gamma^D_S}{\mathcal{H}z}(Y_S-Y_S^{\rm eq})+\epsilon_2\frac{\Gamma^{D}_2}{\mathcal{H}z}(Y_{N_2}-Y_{N_2}^{\rm eq})+
	\epsilon_3\frac{\Gamma^{D}_3}{\mathcal{H}z}(Y_{N_3}-Y_{N_3}^{\rm eq})\nonumber\\&-&\bigg(\frac{1}{2}\frac{\Gamma^{ID}_S+\Gamma^{ID}_{N_2}+\Gamma^{ID}_{N_3}}{\mathcal{H}z}+\frac{\Gamma^{W}_{2}}{\mathcal{H}z} +\frac{\Gamma^{W_1}_{S}+\Gamma^{W_1}_{N_2}+\Gamma^{W_1}_{N_3}}{\mathcal{H}z}  \bigg)Y_{\Delta L}, \nonumber\\\label{eq:dYL2}
\end{eqnarray}}
where $z=M_{S}/T$, $Y_x$ is the abundance of $x$ species defined as, $Y_x=n_x/n_\gamma$, $n_x$ is the number density of $x$, and, {$\mathcal{H}$} is the Hubble parameter. The evolution of $N_2,N_3$ and $S$ are calculated by solving Eqs \ref{eq:dYn2}, \ref{eq:dYn3}, \ref{eq:dYs2}. The Eq \ref{eq:dYL2} gives a net lepton asymmetry generated via the decay of $N_2,N_3$, and $S$.
 \begin{table*}[!]
	\centering
	\caption{Benchmark points for leptogenesis.}
	\resizebox{18cm}{!}{
		\begin{tblr}{
				colspec={llllllll},
				row{1}={font=\bfseries},
				column{1}={font=\itshape},
				row{1}={bg=gray!50},row{2}={bg=gray!40},row{3}={bg=gray!30},row{4}={bg=gray!20},row{5}={bg=gray!10},row{6}={bg=gray!5}
			}
			\toprule BPs&$M_{S}\rm(GeV)$&$M_2/M_S$&$M_3/M_2$&$z_a$&$(y^\dagger_{_{N l}}y_{_{N l}})_{22}$& $y_{_{N S}}$&$\mu_{\rho} \rm(GeV)$ &$T_C \rm(GeV)$&$\lambda_\rho$\\
			\toprule
			BP1&
			$2\times10^{9}$&50.93&322.98&$0-i0.542$& $1.65\times10^{-4}$& $0.35$&$10^9$ &$2.62\times10^9$&$1.5\times10^{-5}$ \\
			
			BP2&
			$2\times10^{9}$&13.19&300.06&$6.42\times10^{-5}-i0.539$& $4.21\times10^{-5}$& $0.13$&$10^9$ &$2.85\times10^9$&$1.5\times10^{-5}$ \\
			
			BP3&
			$4.1\times10^{8}$&18.84&13.27&$0-i0.568$& $1.43\times10^{-5}$& $0.003$&$10^6$ &$3.47\times10^6$&$10^{-3}$\\
			
			BP4&
			$2.8\times10^{6}$&2775.5&22.3&$0.402-i1.955$& $3.309\times10^{-2}$& $0.015$&$10^6$ &$3.47\times10^6$&$10^{-3}$\\
			\bottomrule
	\end{tblr}}
	\label{tab:tab2}
\end{table*}
\begin{figure*}[tbh]
	\includegraphics[scale=0.35]{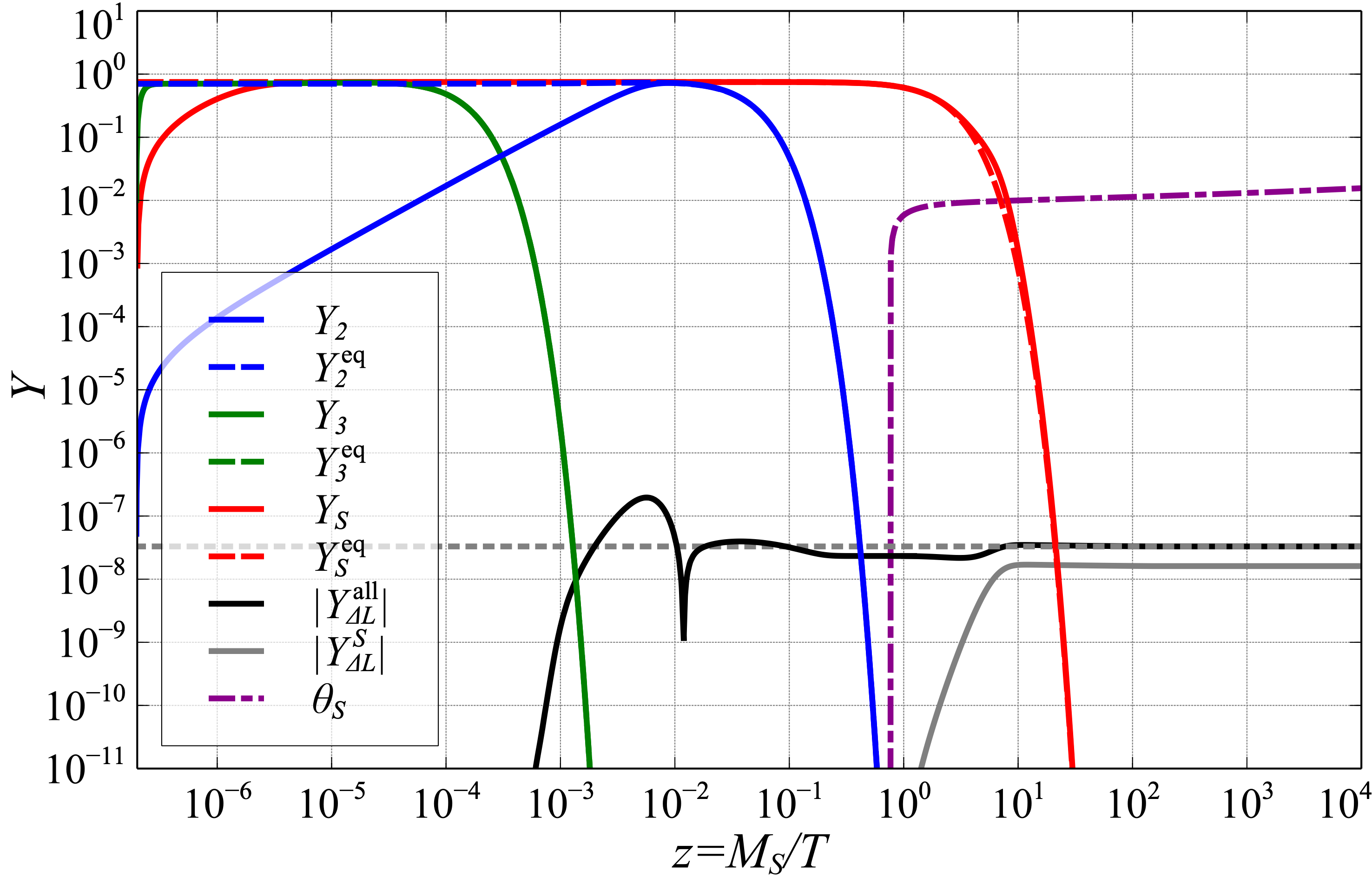}
	\includegraphics[scale=0.35]{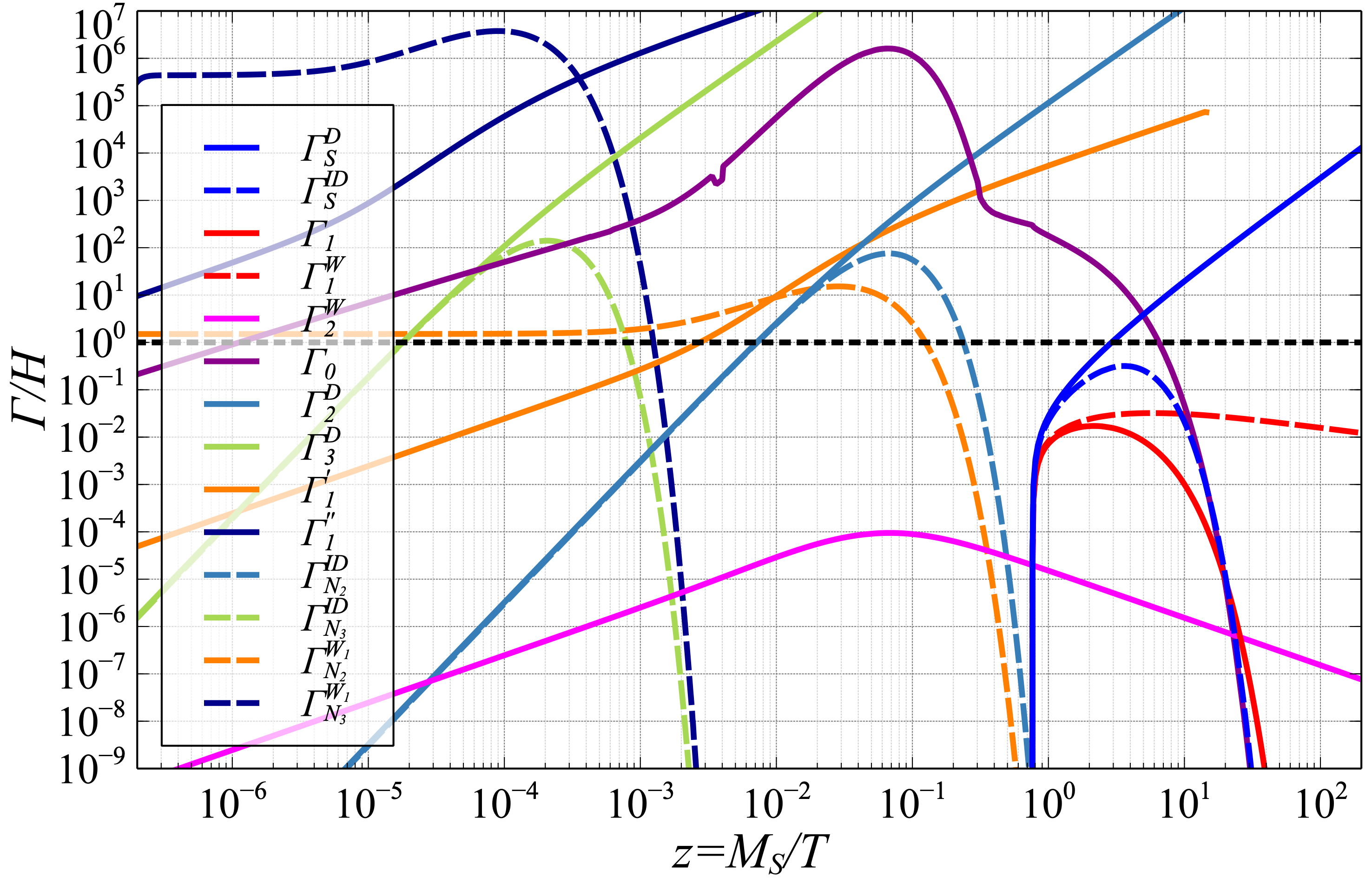}
	\caption{[\textit{left}]Cosmological evolution of the lepton asymmetry and abundances for \textit{BP1} from Table \ref{tab:tab2}. The solid (dashed)
		blue, green, red lines correspond to the abundances (equilibrium abundances) of $N_2,N_3$, and $S$ respectively. The magenta dashed-dotted line represents the mixing angle. The total lepton asymmetry generated via the decay of $N_2,N_3$, and $S$ is shown with black solid line. The gray solid line depicts the lepton asymmetry produced by $S$ only. The horizontal gray dotted line represents the correct lepton asymmetry value. [\textit{right}] The comparison of interaction rates of all the processes involved are shown with respect to the Hubble parameter for the same \textit{BP1}. }\label{fig:ev1}
\end{figure*}
In the above equations, $\Gamma^D_{2},\Gamma^D_{3},\Gamma^D_{S}$ represent the thermal averaged decay width of $N_{2},N_3$, and $S$ respectively. The inverse decay terms due to $N_2,N_3,S$ are $\Gamma^{ID}_{N_2},\Gamma^{ID}_{N_3},\Gamma^{ID}_{S}$. The $\Delta L=1$ scattering interactions due to $N_2,N_3$ and $S$ are encoded in $\Gamma^\prime_1,\Gamma^{\prime\prime}_1$, and $\Gamma_1$, whereas the washout due to $\Delta L=1$ scatterings are given by $\Gamma^{W_1}_{N_2}, \Gamma^{W_1}_{N_3}, \Gamma^{W_1}_{S}$ for $N_2,N_3,S$ in Eq \ref{eq:dYL2}. $\Gamma^W_2$ represents the $\Delta L=2$ washout processes. Here, $\Gamma_{0}$ denotes the $\Delta L=0$ scattering processes like $S\rho\rightarrow\rho S$ in the $t$ and  $s$ channel, and $\Gamma^\prime_{0}$ represents the $\Delta L=0$ scattering processes where two $S$ are in the initial state like, $S\bar{S}\rightarrow\rho\rho,HH^\dagger,L\bar{L}$. These lepton number conserving processes bring the $S$ into thermal equilibrium depending on the strength of the coupling $y_{NS}$. We also calculate lepton asymmetry considering only $S$ contribution by solving  Eq \ref{eq:dYs2} and
{\scriptsize\begin{eqnarray}
	\frac{dY^S_{\Delta L}}{dz}=\epsilon_S\frac{\Gamma_D}{\mathcal{H}z}(Y_S-Y_S^{\rm eq})-\bigg(\frac{1}{2}\frac{\Gamma_{ID}}{\mathcal{H}z}+\frac{\Gamma^W_{1}+\Gamma^W_{2}}{\mathcal{H}z}\bigg)Y_{\Delta L}.\label{eq:dYL}
\end{eqnarray}}
\begin{figure*}[tbh]
	\includegraphics[scale=0.35]{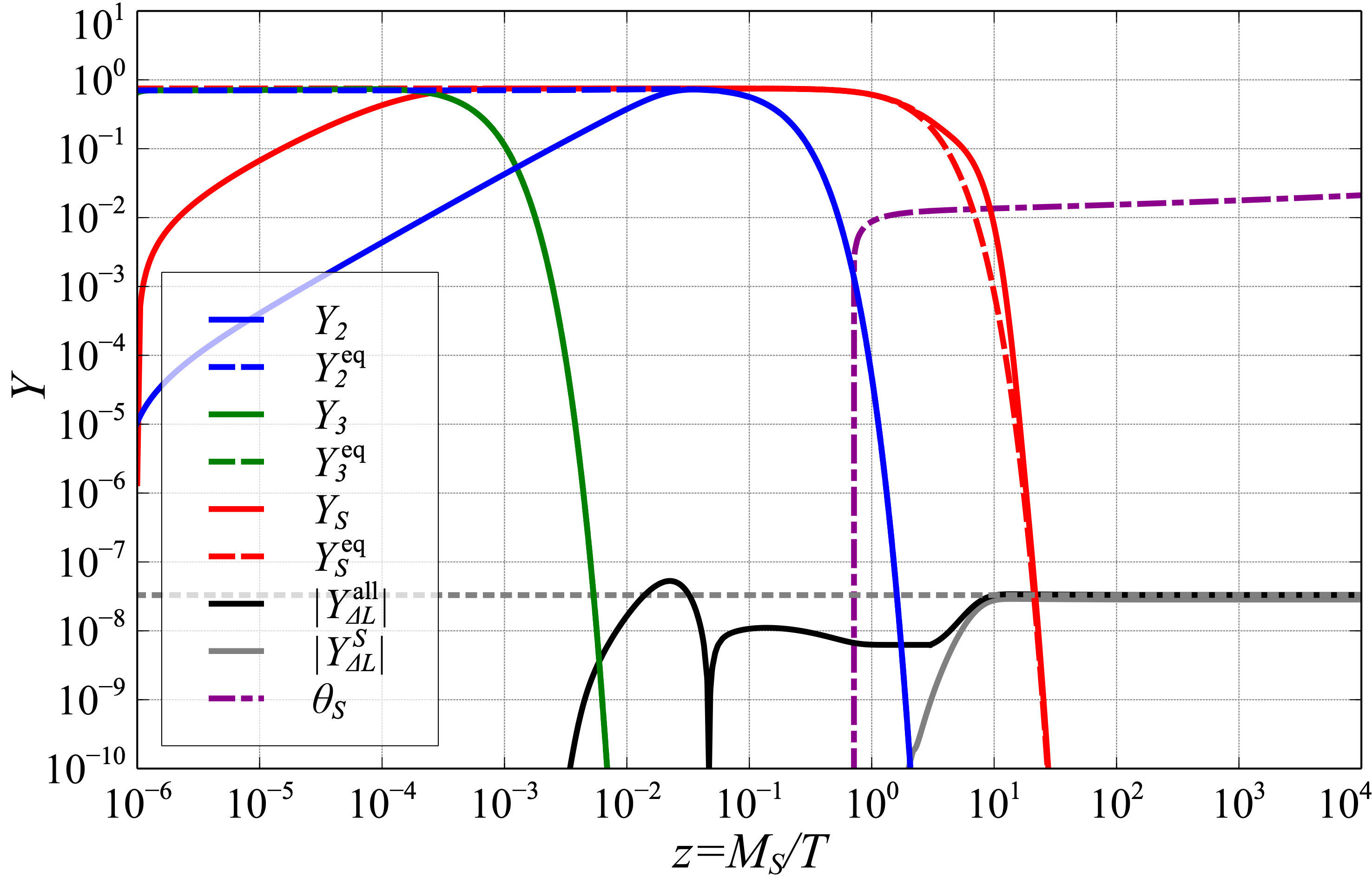}
	\includegraphics[scale=0.35]{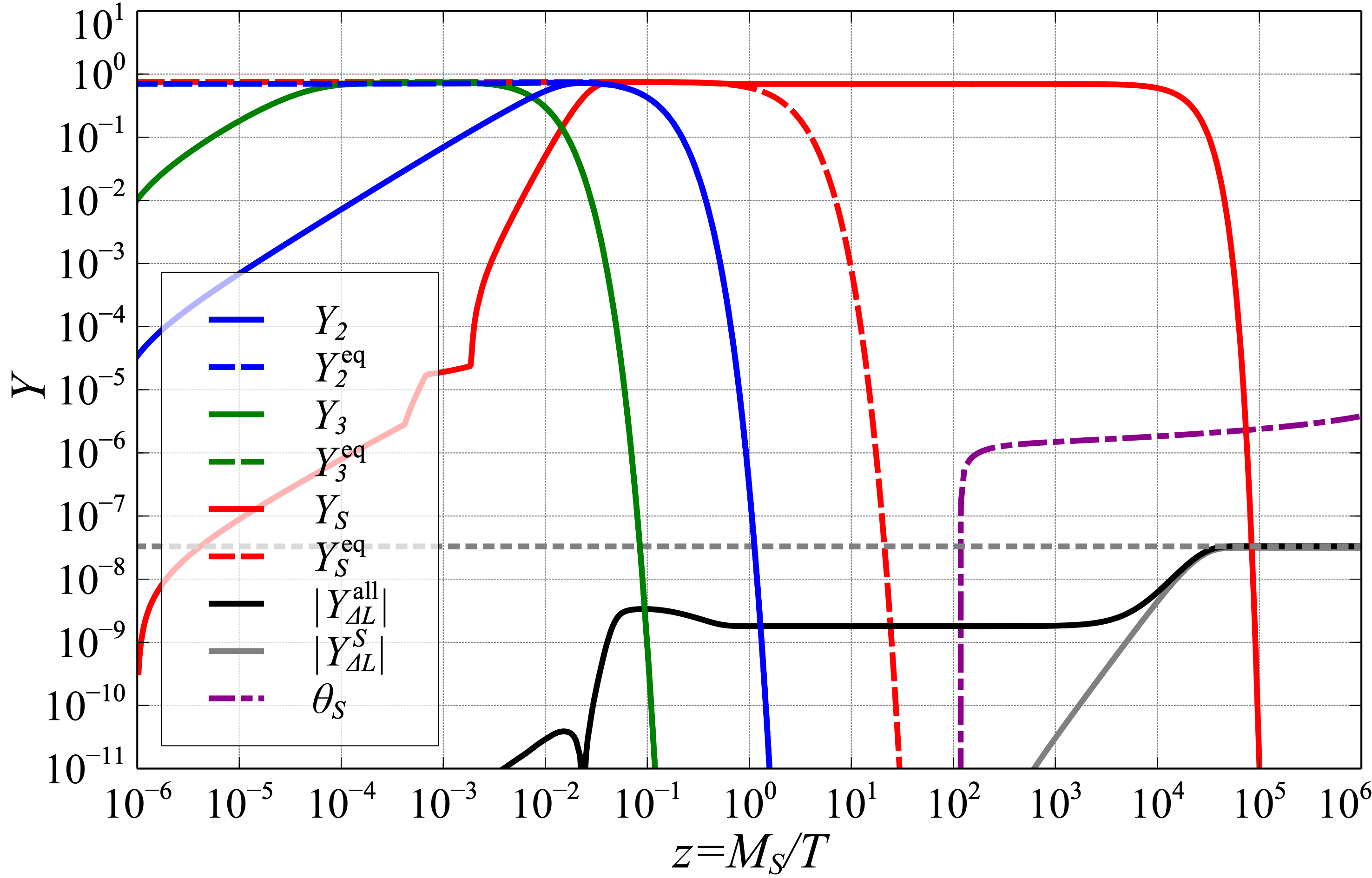}
	\caption{[\textit{left}]Cosmological evolution of the lepton asymmetry and abundances for \textit{BP2} from Table \ref{tab:tab2}. [\textit{right}] Cosmological evolution of the lepton asymmetry and abundances for \textit{BP3} from Table \ref{tab:tab2}. The color code remains same as in Fig \ref{fig:ev1}.}\label{fig:ev23}
\end{figure*}
\begin{figure*}[tbh]
	\includegraphics[scale=0.35]{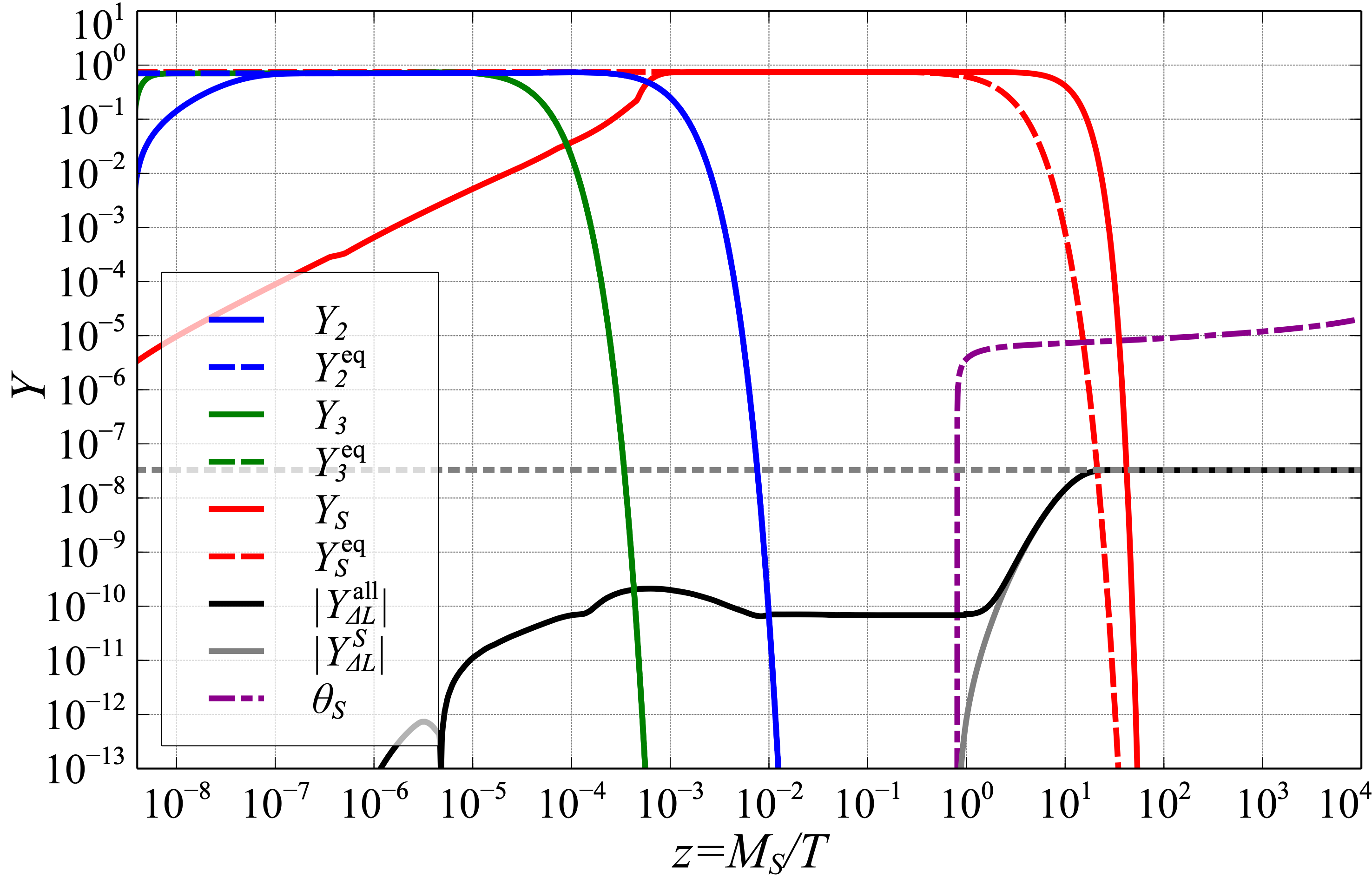}
	\includegraphics[scale=0.35]{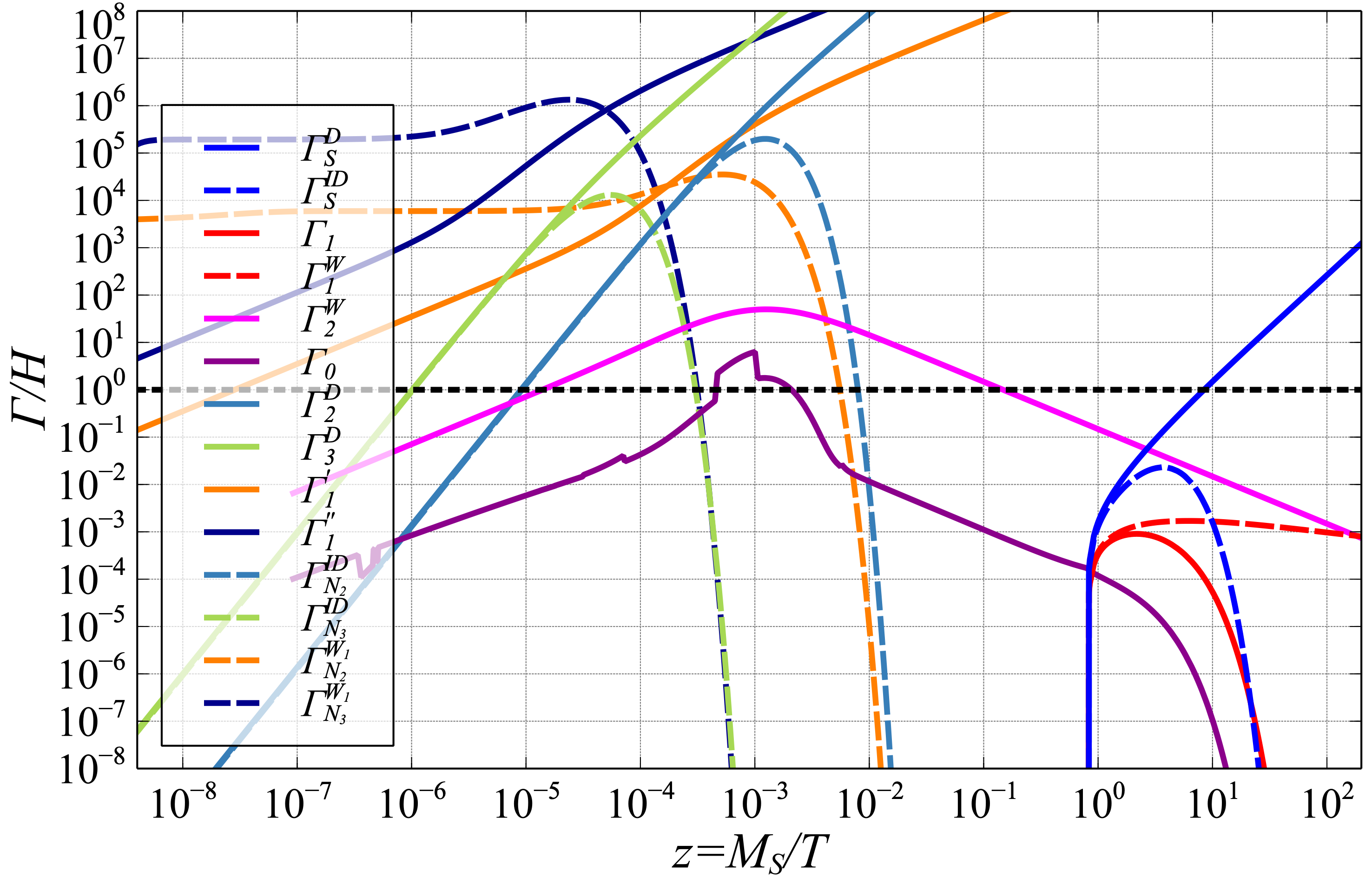}
	\caption{[\textit{left}] Cosmological evolution of the lepton asymmetry and abundances for \textit{BP4} from Table \ref{tab:tab2}. [\textit{right}] The comparison of interaction rates of all the processes involved are shown with respect to the Hubble parameter for the same \textit{BP4}.}\label{fig:ev4}
\end{figure*}

We solve the above equations simultaneously for four sets of benchmark points\footnote{We fix $\lambda_{H\rho}=0.5$ ensuring the stability of the potential.} as mentioned in Table \ref{tab:tab2} to get the lepton asymmetry. Here to evaluate the Yukawa couplings we used the Casas-Ibarra (CI) parametrization as mentioned in Equation \ref{eq:CI} \footnote{For evaluating the Yukawa matrix using the CI parametrization, we use the best fit values of the neutrino oscillation parameters\cite{deSalas:2020pgw}. In Eq \ref{eq:CI} we use the $R$ matrix as $R_{12}R_{23}R_{13}$ with rotation angle $z_a$.}.

\textbf{\underline{\textit{BP1}}:} We first calculate the lepton asymmetry for \textit{BP1} from Table \ref{tab:tab2}, considering only the effect of $S$, \textit{i.e.}, by solving Eqs. \ref{eq:dYs2} and \ref{eq:dYL}, as shown in the \textit{left} panel of Fig. \ref{fig:ev1}. For the chosen parameters, we find the critical temperature to be $T_C = 2.62\times10^{9}$ GeV, \textit{i.e.}, above $2.62\times10^{9}$ GeV the vev of $\rho$ is zero. This implies that the mixing is absent for $z \lesssim 0.76$. In this regime, $S$ reaches equilibrium via $\Delta L = 0$ scatterings that do not involve mixing. For $z > 0.76$, $S$ starts to decay to $L,H$ through mixing with $N_2$. It should be noted that the mixing angle, $\theta_S(T)$ evolves with time, as evident from Eq. \ref{eq:theta1}. The evolution of $\theta_s(T)$ is shown with a magenta dashed-dotted line.  For \textit{BP1}  the \textit{CP} asymmetry parameters are $\epsilon_S = 1.97\times10^{-7}$, $\epsilon_2 = 10^{-5}$, and $\epsilon_3 = 1.07\times10^{-7}$. The resulting lepton asymmetry from the decay of $S$ is calculated to be $|Y^S_{\Delta L}| = 1.61\times10^{-8}$, shown by the gray solid line. The gray dashed line represents the observed lepton asymmetry.

We now include the effects of $N_2$ and $N_3$ by solving Eqs. \ref{eq:dYn2}, \ref{eq:dYn3}, \ref{eq:dYs2}, and \ref{eq:dYL2} for the same \textit{BP1}. The final asymmetry is shown by the black solid line. We observe that $N_2, N_3$ produce an asymmetry of $|Y_{\Delta L}| \sim 2.33\times10^{-8}$, which acts as an initial condition for the lepton asymmetry production via the decay of $S$. The final asymmetry after $S$ fully decays is $|Y^{\rm all}_{\Delta L}| \sim 3.30\times10^{-8}$. The percentage increase in the final asymmetry compared to the $S$-only contribution is 104.94\%. In this case, $\epsilon_{2}\kappa_2 \sim 3\times10^{-8}$ and $\epsilon_{3}\kappa_3 \ll \epsilon_{2}\kappa_2$. This can be interpreted as an $N_2$-dominated leptogenesis scenario, as the asymmetry produced by $N_3$ is entirely washed out by $N_2$, and $S$ alone fails to generate the correct lepton asymmetry—clearly visible from the \textit{left} panel of Fig. \ref{fig:ev1}. In the \textit{right} panel of Fig. \ref{fig:ev1}, the interaction rates of all relevant processes are shown with respect to the Hubble parameter. The different colored lines represent various interaction rates involved in leptogenesis, as indicated in the figure's inset.
	
\textbf{\underline{\textit{BP2}}:} We then move to \textit{BP2}, where we set the mass ratios as $M_{2}/M_S = 13.19$ and $M_3/M_2 = 3957.09$, and choose the Yukawa coupling value as $y_{NS} = 0.13$. The \textit{CP} asymmetry parameters for this \textit{BP2} are $\epsilon_S = 1.95\times10^{-7}$, $\epsilon_2 = 2.57\times10^{-6}$, and $\epsilon_3 = 2.86\times10^{-8}$. The mixing becomes nonzero after $z \sim 0.7$. The lepton asymmetry is generated from the decay of $S$, and the final asymmetry saturates at $|Y^{S}_{\Delta L}| \sim 2.84\times10^{-8}$. The evolution of the asymmetry is shown in the \textit{left} panel of Fig. \ref{fig:ev23} with a gray solid line. After including the contributions from $N_2, N_3$, the asymmetry once $N_2$ has completely decayed is $6.25\times10^{-9}$ which acts as a initial condition for the lepton asymmetry produce via $S$ decay. The total final asymmetry after the decay of $S$ is $|Y^{\rm all}_{\Delta L}| \sim 3.32\times10^{-8}$. This corresponds to a change of only 16.97\% compared to the $S$-only contribution. Note that in this case, $\epsilon_{2}\kappa_2 \sim 8.33\times10^{-9}$ and $\epsilon_{3}\kappa_3 \ll\epsilon_{2}\kappa_2 $.

\textbf{\underline{\textit{BP3}}:} In the \textit{right} panel of Fig. \ref{fig:ev23}, we present the evolution of the asymmetries corresponding to \textit{BP3} from Table \ref{tab:tab2}. The associated \textit{CP} asymmetry parameters are $\epsilon_S = 4.49\times10^{-8}$, $\epsilon_2 = 8.45\times10^{-7}$, and $\epsilon_3 = 7.86\times10^{-8}$. For this parameter choice, the critical temperature is found to be $3.47\times10^6$ GeV. Initially, $S$ reaches equilibrium through lepton number-conserving scatterings. Beyond $z \sim 119$, mixing with $N_2$ sets in, enabling $S$ to decay into $L H$ and generate a lepton asymmetry, as depicted by the gray solid line. This yields a final asymmetry from $S$ decay alone to be $|Y^{S}_{\Delta L}| \sim 3.14\times10^{-8}$. To include the full dynamics, we next incorporate the effects of $N_2$ and $N_3$ along with $S$, with results shown by the black solid line. The asymmetry left behind after the complete decay of $N_2$ is $1.8\times10^{-9}$, with $\epsilon_{2}\kappa_2 \sim 2.4\times10^{-9}$. Subsequently, as $S$ decays (starting near $z \sim 119$), additional asymmetry is generated via $S-N_2$ mixing. The total final lepton asymmetry from the decays of  $N_2,N_3$ and $S$ is estimated to be $|Y^{\rm all}_{\Delta L}| \sim 3.32\times10^{-8}$. Compared to the $S$-only contribution, this gives rise an enhancement of about 5.7\%.

\textbf{\underline{\textit{BP4}}:} In Fig. \ref{fig:ev4}, we compute the lepton asymmetry for the \textit{BP4}, as listed in Table \ref{tab:tab2}. The \textit{CP} asymmetry parameters for this case are $\epsilon_S = 4.46\times10^{-8}$, $\epsilon_2 = 1.24\times10^{-4}$, and $\epsilon_3 = 2.39\times10^{-4}$. The critical temperature is found to be $3.47\times10^6$ GeV. Consequently, $S$ begins decaying to $L H$ around $z \sim 0.81$, generating a lepton asymmetry of $|Y^{S}_{\Delta L}| \sim 3.25\times10^{-8}$. When the contributions from $N_2$ and $N_3$ are included, the final asymmetry increases by only 0.18\%, indicating that $S$ dominates the generation of lepton asymmetry in this scenario. A comparison of the relevant interaction rates is shown in the \textit{right} panel of Fig. \ref{fig:ev4}.

We now show the parameter space where $S$ is the sole degree of freedom contributing to the lepton asymmetry, in the $\epsilon_S\kappa_S$–$M_S$ plane, as illustrated in Fig. \ref{fig:summary}. The free parameters are scanned over the ranges: $M_{2} \in [2.4\times10^{9}, 10^{14}]$ GeV, $M_{3}/M_2 \in [10, 10^4]$, and $z_a \in [-\pi, \pi] + i [-\pi, \pi]$, subject to the condition $M_{2,3} > M_S$. We identify regions in the parameter space where the contributions from $N_2$ and $N_3$ are negligible. This condition is satisfied when $\epsilon_2 \kappa_2,\epsilon_3\kappa_3  \ll 10^{-8}$. These points are then plotted in the $\epsilon_S \kappa_S$–$M_S$ plane using the Eq \ref{eq:cpasymm}. Each line in Fig. \ref{fig:summary} (distinguished by color shading) corresponds to a specific set of parameters \{$z_a, \epsilon_{2}, \epsilon_3, M_2, M_3$\} for which the lepton asymmetry from $N_2$ and $N_3$ remains negligible. The horizontal dashed lines indicate the observed baryon asymmetry, which translates into a required value of the \textit{CP} asymmetry times efficiency factor as $\epsilon_S \kappa_S = (4.38 \pm 0.22) \times 10^{-8}$. By appropriately choosing the other parameters {$y_{NS}, M_S, \lambda_\rho, \mu_{\rho}, \lambda_{H\rho}$}, this asymmetry can be reproduced solely through the contribution from $S$. To better understand the regime of $S$ dominance, we take one set of parameters as \{$z_a=-3.05\times10^{-6}-i1.91, \epsilon_{2}=2.17\times10^{-4}, \epsilon_3=1.39\times10^{-4}, M_2=2.979\times10^{10}{~\rm GeV}, M_3=2.51\times10^{12}{~\rm GeV}$\} for which $\epsilon_2/M_2=7.27\times10^{-15}$, which is shown with the black dashed line. Now for this set of parameters, the green-shaded triangular region, in the plane of $\epsilon_S\kappa_S-M_S$, will be the $S$ dominated regime, which is shown in the inset figure. In other words, in the green shaded region, the contribution from $N_2,N_3$ are negligible and the final asymmetry will be produced solely due to the decay of $S$.\\
From the plot, we observe that the minimum value of $M_S$ that can account for the observed baryon asymmetry is $M_S \simeq 2.8\times10^6$ GeV. For $M_S < 2.8\times10^6$ GeV, values of $\epsilon_S$ alone are insufficient to generate the correct lepton asymmetry. Notably, this lower bound on $M_S$ is \textit{3} orders of magnitude below the Davidson–Ibarra (DI) bound for the scale of thermal leptogenesis, consistent with Eq. \ref{eq:msbound}.

\section{Domain walls and signatures of Gravitational waves from $\mathbf{Z_2}$ symmetry breaking}\label{sec:dws}
The spontaneous breaking of $Z_2$ symmetry giving rise to lepton asymmetry also leads to formation of  domain walls (DWs) in the early Universe. The energy density of the DWs falls with the cosmological scale factor as $\sim R^{-1}$, which is much slower than the matter ($\sim R^{-3}$) and radiation ($\sim R^{-4}$). Thus, the DWs may over close the Universe if they are stable.  This problem can be solved by making the DW unstable and it will disappear eventually in the early Universe. 
 
We demonstrate by considering the potential for the scalar fields as
\begin{eqnarray}
	V(\rho)=-\frac{\mu^2_\rho}{2}\rho^2+\frac{\lambda_\rho}{4}\rho^4+\frac{cT^2}{2}\rho^2=\frac{\lambda_\rho}{4}(\rho^2-v_\rho(T)^2)^2.\nonumber\\\label{eq:rhopot}
\end{eqnarray}
The potential has two degenerate minima at $<\rho>=\pm v_{\rho}(T)$.  The field can occupy any one of the two minima after the symmetry breaking resulting in two different domains, separated by a wall. We consider a static planar DW perpendicular to the $x$ axis in the Minkowski space, $\rho=\rho(x)$. 

The equation of motion for the DW is \cite{Vilenkin:2000jqa,Gelmini:1988sf,Larsson:1996sp,Saikawa:2017hiv,Nakayama:2016gxi},
\begin{eqnarray}
\frac{d^2\rho}{dx^2}-\frac{d V}{d \rho}=0,
\end{eqnarray}
with the boundary condition
\begin{eqnarray}
\lim_{x\rightarrow\pm\infty}\rho(x)=\pm v_\rho.
\end{eqnarray}

After solving the equation of motion, we obtain
\begin{eqnarray}
\rho(x)=v_\rho\tanh(\alpha x),
\end{eqnarray}
where $\alpha\simeq\sqrt{\frac{\lambda_\rho}{2}}v_\rho$.

The DW is extended along $x=0$ plane and the two vacua are realized at $x\rightarrow\pm\infty$. The width of the DW is estimated as $\delta\sim\big(\frac{\sqrt{\lambda_\rho}v_\rho}{\sqrt{2}}\big)^{-1}$. The surface energy density, also referred as tension of the DWs, is calculated to be ,
\begin{eqnarray}
\sigma(T)=\frac{4}{3}\sqrt{\frac{\lambda_\rho}{2}}v_\rho(T)^3\simeq\frac{2}{3}M_{\rho}(T)v_{\rho}(T)^2,
\end{eqnarray}
where $M_{\rho}(T)=\sqrt{2\lambda_\rho}v_\rho(T)$.

As discussed earlier without a soft $Z_2$ breaking term, the DW will be stable and will over close the energy density of the Universe.
In order to over come this problem, we introduce an energy bias in the potential as $\mu^3_{b}\rho/\sqrt{2}$, which breaks the $Z_2$ symmetry explicitly. Here $\mu_{b}$ is a mass dimension one coupling. Equation \ref{eq:rhopot} then becomes
\begin{eqnarray}
\mathcal{V}=V(\rho)+\mu^3_{b}\rho/\sqrt{2}.
\end{eqnarray}
 As a result the degeneracy of the minima is lifted by
\begin{eqnarray}
V_{\rm bias}\equiv |\mathcal{V}(-v_\rho)-\mathcal{V}(v_\rho)|=\sqrt{2}\mu^3_{b}v_\rho(T).
\end{eqnarray}

This creates a pressure difference across the wall\cite{Gelmini:1988sf,Larsson:1996sp,Saikawa:2017hiv}. We assume the annihilation happens in the radiation dominated era. The energy bias has to be large enough, so that the DW can disappear before the BBN epoch, $\it{i.e.}$ $t_{\rm ann}<t_{\rm BBN}$, where
\begin{table*}[!]
	\centering
	\caption{{Benchmark points for gravitational wave}}
	\resizebox{18cm}{!}{
		\begin{tblr}{
				colspec={llllllll},
				row{1}={font=\bfseries},
				column{1}={font=\itshape},
				row{1}={bg=gray!20},row{2}={bg=gray!10},row{3}={bg=gray!5}
			}
			\toprule BPs&$M_S(\rm GeV)$&$M_{2}/M_S$& $T_C(\rm GeV)$&$T_{\rm ann}(\rm GeV)$&$\lambda_\rho(T_{\rm ann})$ &$M_{\rho} (T_{\rm ann}) (\rm GeV)$ & $v_\rho(T_{\rm ann})\rm(GeV)$&$\sigma(\rm TeV^3)$\\
			\toprule
			BPGW1&
			$2\times10^{9}$& $13.19$& $2.85\times10^9$&$5.1\times10^4$&$2.53\times10^{-2}$&$1.42\times10^{9}$&$6.31\times10^{9}$&$3.76\times10^{19}$\\
			
			BPGW2&
			$2.8\times10^{6}$& $277.5$& $3.47\times10^6$&$75$&$1.36\times10^{-4}$&$1.41\times10^{6}$&$8.56\times10^{7}$&$6.91\times10^{12}$\\
			
			\bottomrule
	\end{tblr}}
	\label{tab:tab3}
\end{table*}
\begin{eqnarray}
t_{\rm ann}=\mathcal{C}_{\rm ann}\frac{\mathcal{A}\sigma(T_{\rm ann})}{V_{\rm bias}},
\end{eqnarray} 
where $\mathcal{C}_{\rm ann}$ is a coefficient of $\mathcal{O}(1)$, $\mathcal{A}\simeq0.8\pm0.1$\cite{Hiramatsu:2013qaa} is area parameter, and $t_{\rm BBN}$ is the BBN time scale.
 This gives a lower bound on the $V_{\rm bias}$ as,
\begin{eqnarray}
V_{\rm bias}>6.58\times10^{-14}{\rm GeV^4}\mathcal{C}_{\rm ann}\mathcal{A}\bigg(\frac{10^{-2}{\rm sec}}{t_{\rm BBN}}\bigg)\bigg(\frac{\sigma}{1{\rm TeV^3}}\bigg).\nonumber\\ \label{eq:vbiasbbn}
\end{eqnarray}
\begin{figure*}[t]
	\includegraphics[scale=0.35]{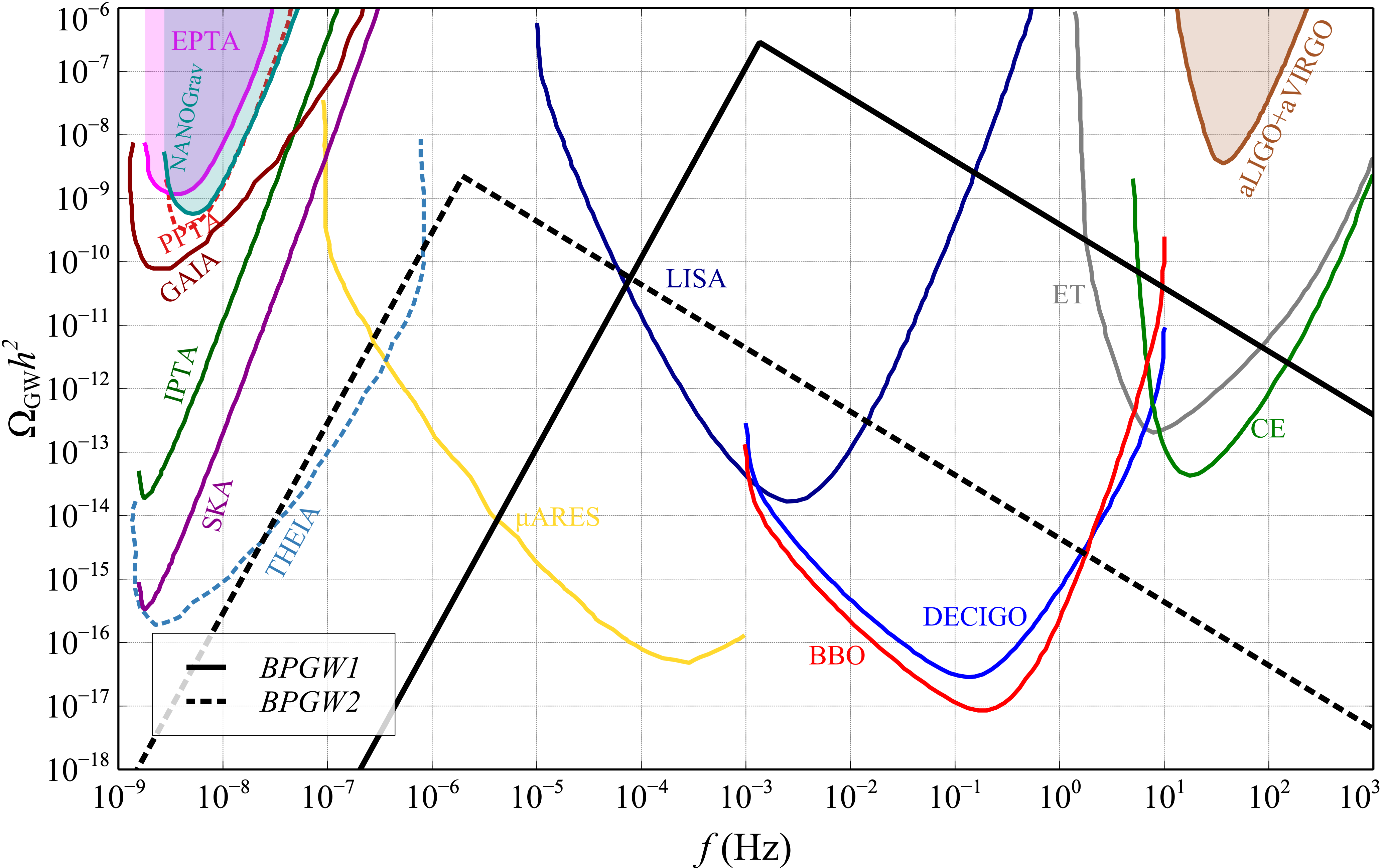}
	\caption{Gravitational wave spectrum from the annihilating domain walls for four benchmark points as mentioned in Table \ref{tab:tab3}. Sensitivities from different gravitational wave search experiments have been shown with different colors.}\label{fig:gw}
\end{figure*}
Eq \ref{eq:vbiasbbn} can be written in terms of the $Z_2$ breaking parameters $\mu_{b}$ as
\begin{eqnarray}
	\mu_{b}&>&1.45839\times10^{-4}{\rm GeV}\mathcal{C}^{\frac{1}{3}}_{\rm ann}\mathcal{A}^{\frac{1}{3}}\bigg(\frac{10^{-2}{\rm sec}}{t_{\rm BBN}}\bigg)^{\frac{1}{3}}\nonumber\\&&\bigg(\frac{M_\rho}{1{\rm TeV}}\bigg)^{\frac{1}{3}}\bigg(\frac{v_\rho}{10^5{\rm TeV}}\bigg)^{\frac{1}{3}}.
\end{eqnarray}

The DWs have to disappear before they could start to dominate the energy density of the Universe, $\it i.e$ $t_{\rm ann}<t_{\rm dom}$, where
\begin{eqnarray}
t_{\rm dom}=\frac{3}{32\pi}\frac{M_{pl}^2}{\mathcal{A}\sigma}.
\end{eqnarray}

This puts a lower bound on the annihilation temperature as,
\begin{eqnarray}
	T_{\rm ann}&>&1.34772{\rm GeV}\mathcal{A}^{1/2}\bigg(\frac{g_*(T_{\rm ann})}{10}\bigg)^{-\frac{1}{4}}\bigg(\frac{v_\rho}{10^5{\rm TeV}}\bigg)\nonumber\\&&\bigg(\frac{M_{\rho}}{{1\rm TeV}}\bigg)^{\frac{1}{2}}.\label{eq:dombound}
\end{eqnarray}

In terms of $\mu_{b}$ it can be expressed as
\begin{eqnarray}
	\mu_{b}&>&4.1404\times10^{-3}{\rm GeV}\mathcal{C}_{\rm ann}^{\frac{1}{3}}\mathcal{A}^{\frac{2}{3}}\bigg(\frac{M_\rho}{1{\rm TeV}}\bigg)^{\frac{2}{3}}\bigg(\frac{v_\rho}{10^5{\rm TeV}}\bigg).\nonumber\\
	\end{eqnarray}
\begin{figure*}[tbh]
	\includegraphics[scale=0.35]{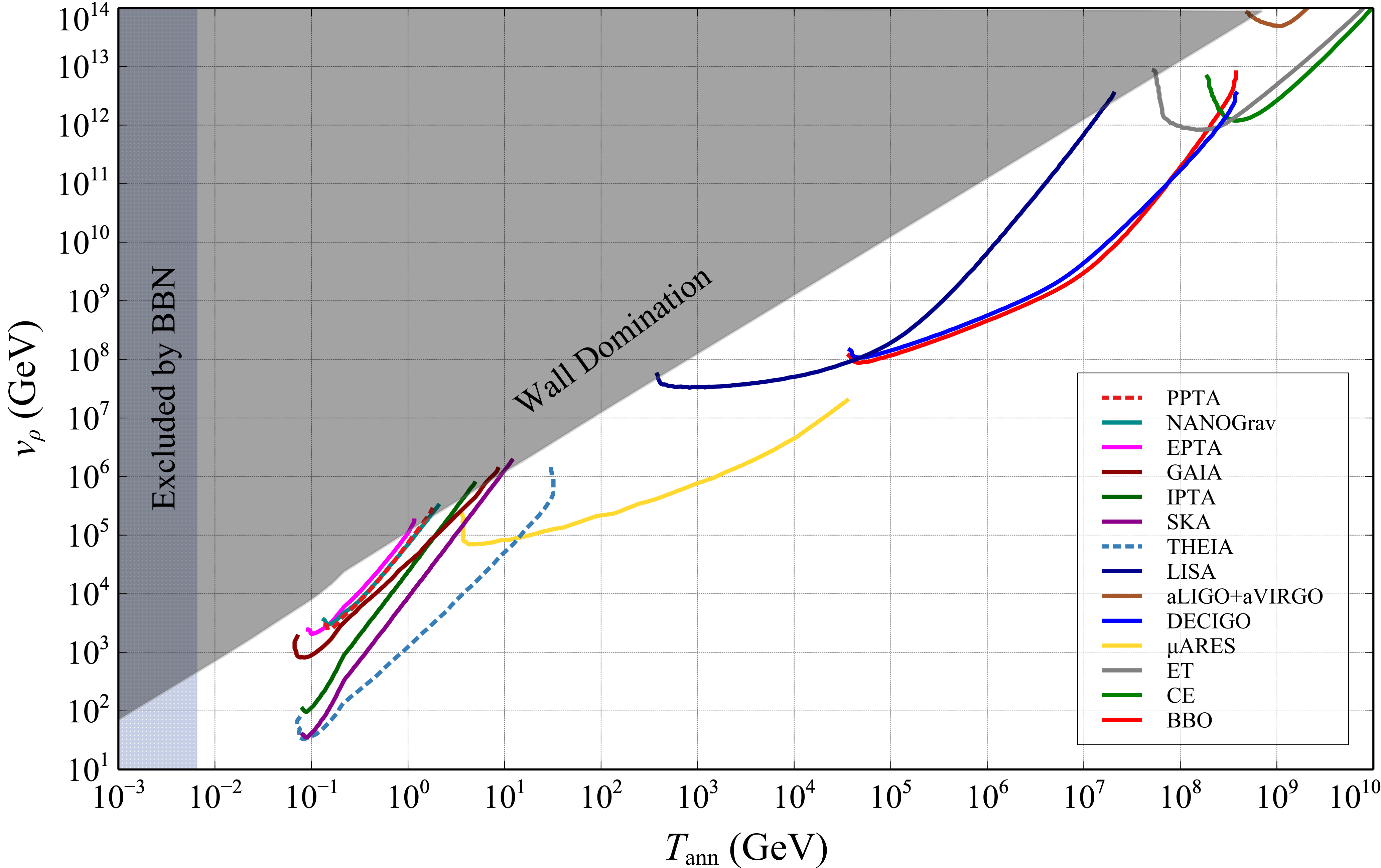}
	\caption{The sensitivity of the gravitational wave search  experiments in the plane of $v_{\rho}-T_{\rm ann}$. These contours are obtained for peak frequency and peak amplitude of the GW as per Eq \ref{eq:peakfre}, and \ref{eq:peakamp} respectively.}\label{fig:gwsummary}
\end{figure*}
 The DWs can then annihilate and emit their energy in the form of stochastic gravitational waves (GWs) which can be detectable at present time.

The peak amplitude of the GW spectrum at the present time, $t_0$, is given by\cite{Saikawa:2017hiv}
\begin{eqnarray}
	\Omega_{\rm GW}h^2(t_0)|_{\rm peak}&=&7.18824\times10^{-18}\mathcal{A}^2\tilde{\epsilon}_{\rm GW} \bigg(\frac{\sigma}{1{\rm TeV^3}}\bigg)^2\nonumber\\&&\bigg(\frac{g_{*s}(T_{\rm ann})}{10}\bigg)^{-\frac{4}{3}}\bigg(\frac{T_{\rm ann}}{10^{-2}\rm GeV}\bigg)^{-4},\label{eq:peakamp}
\end{eqnarray}
where $\tilde{\epsilon}_{\rm GW}\simeq0.7\pm0.4$\cite{Hiramatsu:2013qaa} is the efficiency parameter, $T_{\rm ann}$ is the temperature at which the DWs annihilate, $g_{*s}(T_{\rm ann})$ is the relativistic entropy degrees of freedom at the epoch of DWs annihilation.

From Eq \ref{eq:peakamp}, we see that the peak of GW spectrum is directly proportional to $\sigma^2$ and inversely proportional to $T_{\rm ann}^4$.  In our setup even though DW are produced before EW phase transition, they can sustain until late epoch to give rise larger peak amplitude of the GW spectrum.  If the DW annihilation is happening at an earlier time i.e. at a large temperature, $T_1>T_2$, then the amplitude of the GW will be larger for $T_2$ as compared to $T_1$. 

Assuming the DW disappear at temperature $T_{\rm ann}$, the peak frequency of the GW spectrum at present time is estimated as 
\begin{eqnarray}
	f_{\rm peak}(t_0)&=&1.78648\times10^{-10}{\rm Hz}\bigg(\frac{g_{*s}(T_{\rm ann})}{10}  \bigg)^{-\frac{1}{3}}\nonumber\\&&\bigg(\frac{g_{*}(T_{\rm ann})}{10}\bigg)^{\frac{1}{2}}\bigg(\frac{T_{\rm ann}}{10^{-2}\rm GeV}\bigg).\label{eq:peakfre}
\end{eqnarray}
The amplitude of the GW for any frequency at the present time varies as
\begin{equation}
\Omega_{\rm GW}h^2(t_0,f) =\Omega_{\rm GW}h^2(t_0)|_{\rm peak}\left\{
	\begin{array}{l}
	\frac{f_{\rm peak}}{f}~~~~~~~~~~f>f_{\rm peak}\\
	\bigg(\frac{f}{f_{\rm peak}}\bigg)^3~~~~f<f_{\rm peak}.\\
	\end{array}
	\right.
\end{equation}

 It is worth mentioning that $v_\rho$ is the only parameter that is sensitive to both the leptogenesis and GW spectrum.  In Fig \ref{fig:gw}, we have illustrated the GW spectrum for two benchmarks as mentioned in Table \ref{tab:tab3}. We have shown different sensitivities from experiments BBO\cite{Yunes:2008tw}, CE, DECIGO\cite{Adelberger:2005bt}, NANOGrav\cite{NANOGrav:2023gor,NANOGrav:2023hvm}, EPTA \cite{EPTA:2023fyk}, CPTA \cite{Xu:2023wog} , PPTA \cite{Reardon:2023gzh}, ET\cite{Punturo:2010zz}, GAIA\cite{Garcia-Bellido:2021zgu}, IPTA \cite{Hobbs:2009yy}, LISA,   SKA\cite{Weltman:2018zrl}, THEIA\cite{Garcia-Bellido:2021zgu}, aLIGO \cite{LIGOScientific:2014pky}, aVIRGO, and $\mu$ARES\cite{Sesana:2019vho}. For \textit{BPGW1}, which corresponds to \textit{BP2} in Table \ref{tab:tab2}, the lower bound on the domain wall (DW) annihilation temperature is given by Eq. \ref{eq:dombound} as $5.0171\times10^4$ GeV. We choose $5.1\times10^4$ GeV as the annihilation temperature for this case, leading to a surface energy density of $3.76\times10^{19}~{\rm TeV^3}$ and a GW peak frequency of $\sim$ 0.0013 Hz. This benchmark lies inside the sensitivity range of space-based interferometers such as LISA, DECIGO, as well as ${\rm \mu ARES}$, ET, and CE. Moving to \textit{BPGW2} (which is same as \textit{BP4} in Table \ref{tab:tab2}), the lower bound on the DW annihilation temperature is found to be 72 GeV. We take $75$ GeV as the annihilation temperature, resulting in a surface energy density of $6.91\times10^{12}~{\rm TeV^3}$. Although this energy density is much lower than that of \textit{BPGW1}, the reduced annihilation temperature compensates a little for the decrease in GW amplitude, as described by Eq. \ref{eq:peakamp}. The peak GW frequency is $1.9\times10^{-6}$ Hz, placing this benchmark inside the sensitivity range of experiments such as THEIA, ${\rm \mu{ARES}}$, LISA, BBO, and DECIGO. At this juncture, we note that the recent study, \cite{Bagherian:2025puf} has pointed out that BBN can be sensitive to domain wall annihilation occurring near the BBN epoch (MeV scale). However, in our scenario, the domain wall annihilation takes place at $T_{\rm ann}=$75 GeV, which is well before the onset of BBN, making it safe from such constraints.
  
Figure \ref{fig:gwsummary} illustrates the sensitivity reaches of various experiments in the $v_\rho$ versus $T_{\rm ann}$ plane corresponding to \textit{BPGW1} in Table \ref{tab:tab3}. Here we treat the vev of $\rho$ as independent parameter. The parameter space is constrained by two key requirements: (i) domain walls (DWs) must annihilate before big bang nucleosynthesis, and (ii) they must do so before dominating the energy density of the Universe.

\section{Dark Matter phenomenology}\label{sec:dmpheno}
\begin{figure*}[!]
	\includegraphics[scale=0.35]{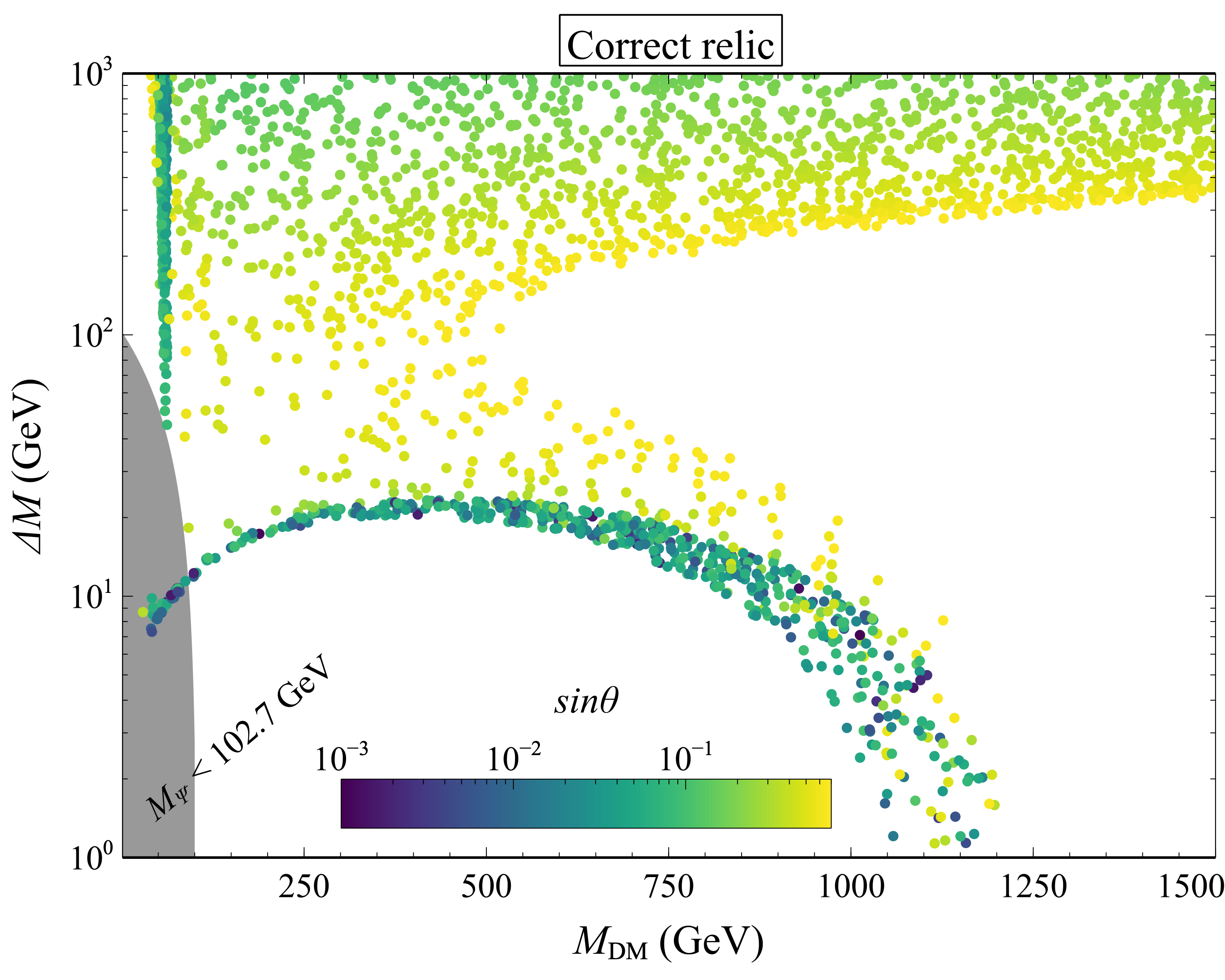}
	\includegraphics[scale=0.35]{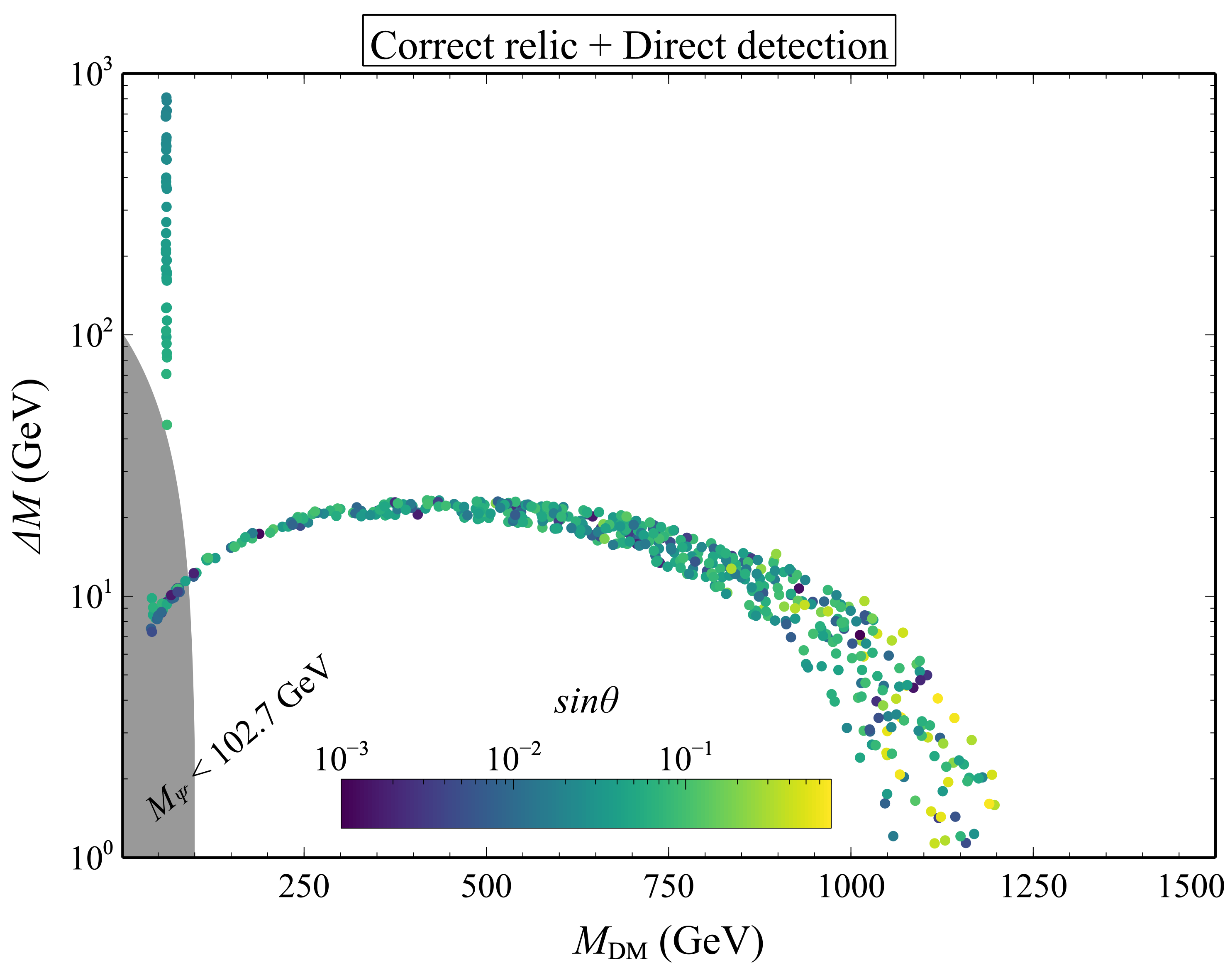}
	\caption{[\textit{left}]: DM parameter space satisfying correct relic in the plane of $\Delta M$ vs $M_{\rm DM}$.  [\textit{right}]: Parameter space consistent with both relic density and direct search constraints	(LZ) in the $\Delta M$ versus $M_{\rm DM}$  plane. The gray shaded region in the bottom left corner is ruled out by the LEP
		exclusion bound on charged fermion mass $M_{\Psi^+}\equiv M>$ 102.7 GeV.}
	\label{fig:dm}
\end{figure*}
We turn to comment on DM in our setup. Due to the unbroken $Z_2^\prime$ symmetry, combination of $N_1$ and $\Psi=(\psi^0 ~~\psi^-)^T\equiv(\psi_L^0+\psi_R^0 ~~\psi^-)^T$ can give rise to a singlet-doublet Majorana DM\cite{Dutta:2020xwn}. The relevant DM Lagrangian reads as
\begin{eqnarray}
	\mathcal{L}_{\rm DM}&=&\overline{\Psi}i\gamma^\mu\mathcal{D}_\mu\Psi-M\bar{\Psi}\Psi+\overline{N_1}i\gamma^\mu\partial_\mu N_1-\frac{1}{2}M_1\overline{N^c_1}N_1
	\nonumber\\&-&\frac{y_1}{\sqrt{2}}\overline{\Psi}\tilde{H}(N_1+N^c_1)+h.c..\nonumber\\
\end{eqnarray}
The neutral fermion mass matrix can be written in the basis $((\psi^0_R)^c,\psi^0_L,(N_1)^c)^T$ as
\begin{equation}
\begin{pmatrix}
	0 & M & \frac{m_D}{\sqrt{2}}\\
	M&0&  \frac{m_D}{\sqrt{2}}\\
	 \frac{m_D}{\sqrt{2}} &  \frac{m_D}{\sqrt{2}}& M_1
\end{pmatrix},
\end{equation}
where $m_D=y_1 v_h/\sqrt{2}$.
The mass matrix can be diagonalized with a unitary matrix of the form $U(\theta)=U_{13}(\theta_{13}=\theta).U_{23}(\theta_{23}=0).U_{12}(\theta_{12}=\frac{\pi}{4})$.
The three neutral states mix and gives three Majorana states as $\chi_i=\frac{\chi_{iL}+\chi_{iL}^c}{\sqrt{2}}$, where
\begin{eqnarray*}
\chi_{1L}=\frac{\cos\theta}{\sqrt{2}}(\psi^0_L+(\psi^0_R)^c)+\sin\theta N_1^c,
\end{eqnarray*}
\begin{eqnarray*}
\chi_{2L}=\frac{i}{\sqrt{2}}(\psi^0_L-(\psi^0_R)^c),
\end{eqnarray*}
\begin{eqnarray}
\chi_{3L}=-\frac{\sin\theta}{\sqrt{2}}(\psi^0_L+(\psi^0_R)^c)+\cos\theta N_1^c.
\end{eqnarray}
The corresponding mass eigenvalues are
\begin{eqnarray}
M_{\chi_1}&=&M\cos^2\theta+M_1\sin^2\theta+m_D\sin2\theta,\nonumber\\M_{\chi_2}&=&M,\nonumber\\
M_{\chi_3}&=&M\sin^2\theta+M_1\cos^2\theta-m_D\sin2\theta,
\end{eqnarray}
where the mixing angle is given as
\begin{eqnarray}
	\tan2\theta=\frac{2m_D}{M-M_1}.
\end{eqnarray}
Here we identify the $\chi_3$ be DM candidate. The Yukawa coupling can be expressed as,
\begin{eqnarray}
y_1=\frac{{\Delta}M\sin2\theta}{\sqrt{2}v_h},
\end{eqnarray}
where ${\Delta}M$ is the mass splitting between DM and the next heavy neutral fermion state. The free parameters in the DM phenomenology are $\{M_{\chi_3}\equiv M_{{\rm DM}},{\Delta}M=M_{\chi_1}-M_{\chi_3}\approx M_{\chi_2}-M_{\chi_3},\sin\theta\}$.

The DM relic is decided by the freeze-out of various annihilation and coannihilation processes  in the early Universe. We compute the relic density and the DM spin-independent cross-section using the micrOMEGAs package \cite{Alguero:2023zol}.  In Fig \ref{fig:dm} [\textit{left}], we show the mass splitting between the DM and the light neutral fermion as a function of DM mass for correct relic density. As the DM mass increases, the annihilation cross-section decreases, and as a result the relic increases. The coannihilation plays important role in bringing down the relic to correct ball park. The coannihilation is large when the mass splitting between the DM and the next $Z_2^\prime$ odd sector particle is smaller. This feature is clearly visible in Fig \ref{fig:dm} [\textit{left}]. When the mass splitting $\Delta M$ is large, coannihilation becomes negligible. The Higgs mediated annihilation processes decide the relic density mainly. Thus the mixing angle $\sin\theta$ becomes important here as the Yukawa coupling is $\propto\Delta M\sin2\theta$. The dependence of mixing angle for large mass splitting is prominent in Fig \ref{fig:dm} [\textit{left}]. For fixed $\sin\theta$, as the mass splitting increases, the coupling increases, which leads to an increase in annihilation cross-section, and thus, the relic decreases. To get the correct relic, the DM mass has to be large, which makes the cross-section smaller giving the correct relic. Larger $\sin\theta$ requires smaller $\Delta M$ to give the correct relic. 
We take these correct relic data points and impose the direct detection constraint from LZ experiment \cite{LZ:2022lsv} and show in Fig \ref{fig:dm} [\textit{right}]. The direct detection is possible via the Higgs portal. The spin-independent DM-nucleon cross-section is $\sigma^{SI}_{{\rm DM-N}}\propto y^2_1\sin^22\theta\propto \Delta M^2\sin^42\theta$. A larger mixing angle as well as larger mass splitting will result in larger direct detection cross-section. However, larger mass splitting is allowed for DM mass at Higgs resonance. We also impose the LEP bound on the mass of the charged  component of the doublet, which is $M_{\Psi^+}>102.7$ GeV \cite{DELPHI:2003uqw}.
\section{Conclusion}\label{sec:conclusion}
In this paper, we explored the potential for generating successful thermal leptogenesis at a scale lower than the Davidson Ibarra bound in an extended type-I seesaw framework along with nonzero neutrino mass, dark matter and gravitational wave. In our setup the lightest RHN $N_1$ mixes with the neutral component of a vectorlike fermion doublet $\Psi$ which are odd under an unbroken $Z_2^\prime $ symmetry. As a result we get singlet-doublet Majorana DM in a wide range of parameter space. We also added a singlet scalar $\rho$ and a singlet fermion $S$ in the canonical type-I seesaw which are odd under an imposed discrete $Z_2$ symmetry. At high scale, typically above the EWPT $\rho$ acquires a vev and breaks the $Z_2$ symmetry spontaneously. As a result we got $S$ mixed up with $N_{2}$ such that late decay of $S$ could give rise to a relatively low scale thermal leptogenesis. In particular we saw that successful thermal leptogenesis requires $M_S\gtrsim 2.8\times10^6$ GeV.  We note that this is \textit{3} orders magnitude smaller than the usual Davidson Ibarra bound ($M_N>2.4\times10^9$ GeV). The spontaneous breaking of the $Z_2$ symmetry also gave rise to DWs in the early Universe. We discussed the evolution of DWs, which annihilate by emitting stochastic GWs. In the appropriate parameter space, these GWs can produce detectable signatures at space-based interferometers such as LISA, UDECIGO, BBO as well as ET, CE, THEIA,SKA, ${\rm \mu ARES}$. In our scenario, the GW frequencies lie in the nHz to kHz range, making them relevant for high-frequency GW detectors as well as low-frequency detectors.

The scenario we discussed here, can be implemented  in a gauged $U(1)_{\rm B-L}$ symmetric model where the connection between the DM, leptogenesis and GW is more coherent. In this case, the presence of three RHNs, $N_{1,2,3}$ is required to cancel the gauge anomaly. The minimal $U(1)_{\rm B-L}$ model can be extended with a vectorlike singlet fermion $S$, a vectorlike doublet fermion $\Psi$, and one singlet scalar $\rho$. Additionally a $Z_2$ symmetry is imposed under which $S$, and $\rho$ are odd, while all other particles are even. The lightest RHN, $N_1$ along with $\Psi$ can be made a viable DM candidate by assigning another discrete symmetry $Z^\prime_{2}$ under which they are odd and all the other particles are even. All the fermions ($N_{1,2,3},S$) will get their masses from the $U(1)_{\rm B-L}$ breaking scalar $\Phi$. The leptogenesis will be generated by the decay of $S$ while the neutrino mass can be realized via the type-I seesaw mechanism anchored by $N_{2,3}$. In this scenario there will be two sources of GWs: (i) the breaking of $U(1)_{\rm B-L}$ symmetry, and (ii) the annihilation of DWs that arise due to breaking of the $Z_2$ symmetry by the vev of $\rho$.

\noindent
\acknowledgments	

P.K.P. would like to acknowledge the Ministry of Education, Government of India, for providing financial support for his research via the Prime Minister’s Research Fellowship (PMRF) scheme. The work of N.S. and P.S. is supported by the Department of Atomic Energy-Board of Research in Nuclear Sciences, Government of India (Ref. Number: 58/14/15/2021- BRNS/37220). P.K.P would like to thank Satyabrata Mahapatra and Pram Milan P Robin for useful discussions.

\appendix
\section{Lower bound on the  leptogenesis scale in canonical type-I seesaw}\label{sec:canocialtype1lower}
In the hierarchical scenario of the canonical type-I leptogenesis the lightest RHN decays to $L,H$ and $\bar{L},H^{\dagger}$. The interference between these tree level and one loop processes can give rise to a \textit{CP} asymmetry given as 
\begin{eqnarray}
	\epsilon_{_{N_1}}=-\frac{3}{8\pi v_h^2}M_{1}\sum_{j=2,3}\frac{1}{M_j}\frac{Im[(m^\dagger_Dm_D)_{1j}]^2}{(m^\dagger_Dm_D)_{11}},\label{eq:CPasytype3}
\end{eqnarray}
where $v_h=246$ GeV, is the SM Higgs vev.

The decay width of $N_1$ is calculated to be,
\begin{eqnarray}
	\Gamma_{_{N_1}}=\frac{(m^\dagger_{D}m_{D})_{11}}{8\pi v_h^2}M_{{1}}.\label{eq:gammatype1}
\end{eqnarray}
A net lepton asymmetry can be generated once this decay rate falls below the Hubble expansion rate of the Universe,
\begin{equation}
	\mathcal{H} (T)=1.66\sqrt{g_*}\frac{T^2}{M_{pl}},
\end{equation}
where $g_*$ is the effective number of relativistic degrees of freedom, $M_{pl}=1.22\times10^{19}$ GeV is the Planck mass.

The \textit{CP} asymmetry is bounded from above as \cite{Davidson:2002qv}
\begin{equation}
	\epsilon_{_{N_1}}\leq\frac{3}{8\pi v_h^2}M_{1} \sqrt{\Delta m^2_{atm}}.
	\label{eq:CPasytype31}
\end{equation} 
In the zero initial abundance case the lower bound on the lightest RHN mass is found to be \cite{Buchmuller:2002rq}
\begin{equation}
	M_{1}\gtrsim2.4\times10^{9}~ {\rm GeV}.
\end{equation} 
This is mainly because the same coupling is responsible to give neutrino mass as well as leptogenesis. Various attempts have been made to lower this leptogenesis scale\footnote{The corresponding super symmetric theory requires a low re-heating temperature \cite{Khlopov:1984pf,Ellis:1984eq,Ellis:1984er,Bolz:2000fu,Kawasaki:2004qu,Allahverdi:2005mz} due to the over production of gravitino which is in conflict with the lower bound on the lightest RHN mass $M_{1}\gtrsim2.4\times10^{9}$ GeV.}, \textit{e.g.} incorporating flavor effects \cite{Blanchet:2008pw}, by adding
extra scalar fields \cite{Clarke:2015hta,Hugle:2018qbw,Vatsyayan:2022rth}, resonant leptogenesis \cite{Pilaftsis:2003gt}, and by decoupling neutrino mass and leptogenesis \cite{Ma:2006te,Ma:2006ci}. 

\section{Renormalization Group Equations}\label{app:rgeq}
The RG equations are given as
{\allowdisplaybreaks  \begin{align} 
		\beta_{g_1}^{(1)} & =  
		\frac{41}{10} g_{1}^{3} ,\\\nonumber\\
		\beta_{g_2}^{(1)} & =  
		-\frac{19}{6} g_{2}^{3}, \\ 
		\beta_{g_3}^{(1)} & =  
		-7 g_{3}^{3},\\
		\beta_{\lambda_\rho}^{(1)} & =  
		2 \Big(4 \lambda_\rho y^2_{NS}  -4 y^4_{NS}  + 9 \lambda_{\rho}^{2}  + \lambda_{H\rho}^{2}\Big),\\		
		\beta_{\lambda_{H\rho}}^{(1)} & = 4 \lambda_{H\rho} y^2_{NS} -8y^2_{NS}y^2_{Nl} +\frac{1}{10} \lambda_{H\rho} \Big(120 \lambda_H+ 20 y^2_{Nl}\nonumber\\&  + 40 \lambda_{H\rho}  -45 g_{2}^{2}  + 60 \lambda_\rho  -9 g_{1}^{2}+60y^2_t \Big),\\ 
		\beta_{\lambda_H}^{(1)} & =  
		\frac{27}{200} g_{1}^{4} +\frac{9}{20} g_{1}^{2} g_{2}^{2} +\frac{9}{8} g_{2}^{4} -\frac{9}{5} g_{1}^{2} \lambda_H -9 g_{2}^{2} \lambda_H +24 \lambda_H^{2}\nonumber\\&+\frac{1}{2} \lambda_{H\rho}^{2} +4 \lambda_H y^2_{Nl} -2y^4_{Nl}+12\lambda_H y^2_t-6y^4_t,\\
		\beta_{y_{NS}}^{(1)} & =5y_{NS}^3+y_{NS}y_{Nl}^2,\\
	\beta_{y_{Nl}}^{(1)} & =\frac{1}{2}\Big( y_{NS}^2y_{Nl}+3y_{Nl}^3 \Big)+y_{Nl}\Big( -\frac{9}{20}g_1^2-\frac{9}{4}g_2^2+y_{Nl}^2+3y_t^2 \Big),\nonumber\\	
	\beta_{y_{t}}^{(1)} & =\frac{3}{2}y^3_t+y_t\Big( -\frac{17}{20}g_1^2-\frac{9}{4}g_2^2-8g_3^2+y_{Nl}^2+3y_t^2 \Big)
\end{align}} 
Here the beta functions are defined as $\beta^{(1)}_{x}=16\pi^2\frac{dx}{d\log\mu}$.

\end{document}